\documentclass[fleqn,usenatbib]{mnras}

\usepackage{newtxtext,newtxmath}
\newcommand{\HAL}{HAL}
\usepackage[T1]{fontenc}
\usepackage{multirow}
\usepackage{multicol}
\usepackage{makecell}
\usepackage{graphicx}
\usepackage[flushleft]{threeparttable}
\usepackage{booktabs}
\DeclareUnicodeCharacter{2212}{-}
\DeclareRobustCommand{\VAN}[3]{#2}
\let\VANthebibliography\thebibliography
\def\thebibliography{\DeclareRobustCommand{\VAN}[3]{##3}\VANthebibliography}

\usepackage{graphicx}	
\usepackage{amsmath}	

\title[Acetaldehyde BE]{Acetaldehyde binding energies: a coupled experimental and theoretical study}

\author[Ferrero et al.]{
S. Ferrero$^{1}$,
F. Grieco$^{2,3}$,
A-S. Ibrahim Mohamed$^{2}$,
F. Dulieu$^{2}$\thanks{E-mail: francois.dulieu@cyu.fr},
A. Rimola$^{1}$\thanks{E-mail: albert.rimola@uab.cat},
C. Ceccarelli$^{4}$\thanks{E-mail: cecilia.ceccarelli@univ-grenoble-alpes.fr},
C. Nervi$^{5}$,
\and M. Minissale$^{6}$,
P. Ugliengo$^{5}$
\\
$^{1}$Departament de Química, Universitat Autonoma de Barcelona, 08193 Bellaterra, Catalonia, Spain\\
$^{2}$CY Cergy Paris Université, Observatoire de Paris, PSL University, Sorbonne Université, CNRS, LERMA, F-95000, Cergy, France\\
$^{3}$ University of Ghent, Department of Physics and Astronomy, Ghent, Belgium  \\
$^{4}$Univ. Grenoble Alpes, CNRS, IPAG, 38000 Grenoble, France\\
$^{5}$Dipartimento di Chimica, Università degli Studi di Torino, via P. Giuria 7, 10125, Torino, Italy\\
$^{6}$ AMU Aix-Marseille University -  PIIM laboratory, CNRS, Marseille, France 
}

\date{Accepted XXX. Received YYY; in original form ZZZ}

\pubyear{2021}

\begin{document}
\label{firstpage}
\pagerange{\pageref{firstpage}--\pageref{lastpage}}
\maketitle

\begin{abstract}
Acetaldehyde is one of the most common and abundant gaseous interstellar complex organic molecules, found in cold and hot regions of the molecular interstellar medium. 
Its presence in the gas-phase depends on the chemical formation and destruction routes, and its binding energy (BE) governs whether acetaldehyde remains frozen onto the interstellar dust grains or not.
In this work, we report a combined study of the acetaldehyde BE obtained via laboratory TPD (Temperature Programmed Desorption) experiments and theoretical quantum chemical computations. 
BEs have been measured and computed as a pure acetaldehyde ice and as mixed with both polycrystalline and amorphous water ice.
Both calculations and experiments found a BE distribution on amorphous solid water that covers the 4000--6000 K range, when a pre-exponential factor of $1.1\times 10^{18}s^{-1}$ is used for the interpretation of the experiments. 
We discuss in detail the importance of using a consistent couple of BE and pre-exponential factor values when comparing experiments and computations, as well as when introducing them in astrochemical models. 
Based on the comparison of the acetaldehyde BEs measured and computed in the present work with those of other species, we predict that acetaldehyde is less volatile than formaldehyde, but much more than water, methanol, ethanol, and formamide.
We discuss the astrochemical implications of our findings and how recent astronomical high spatial resolution observations show a chemical differentiation involving acetaldehyde, which can easily explained as due to the different BEs of the observed molecules.
\end{abstract}


\section{Introduction}

Acetaldehyde (CH$_3$CHO) was one of the first polyatomic molecules discovered in the interstellar medium (ISM) \citep{Gottlieb1973,Fourikis1974}.
It is a rather common interstellar molecule, found in warm and cold environments.
For example, it is abundant in hot cores \citep[e.g.][]{Blake1987,Csengeri2019,Law2021} and hot corinos \citep[e.g.][]{Cazaux2003,Manigand2020,Yang2021,Chahine2022}, protostellar molecular shocks \citep[e.g.][]{Lefloch2017,DeSimone2020_acet,codella2020} and young disks \citep[e.g.][]{Codella2018,Lee2019}.
Against the theoretical expectations, acetaldehyde is also present in cold prestellar cores \citep[e.g.][]{Bacmann2012,vastel2014,Scibelli2020}.
Finally, acetaldehyde is detected in comets \citep[][]{Crovisier2004,LeRoy2015,Biver2021}, where the measured relative abundance with respect to methanol is similar to that found in hot corinos \citep[e.g.][]{Bianchi2019,Drozdovskaya2019}.

Several experimental and theoretical studies have focused on the acetaldehyde formation routes both in the gas-phase and on the grain-surfaces.
\cite{Charnley2004} first proposed that acetaldehyde is synthesised in the gas-phase via the reaction of the ethyl radical (CH$_3$CH$_2$) with atomic oxygen (O).
Subsequently, \cite{Skouteris2018} found that acetaldehyde could be a daughter of ethanol (CH$_3$CH$_2$OH), while \cite{Vasyunin2017} proposed a synthesis from methanol (CH$_3$OH) reacting with CH.
Finally, publicly available astrochemical reaction networks \citep[KIDA and UMIST:][respectively]{Wakelam2012,mcelroy2013} also report an ionic route via the protonated acetaldehyde (CH$_3$CHOH$^+$), which is in turn formed from dimethyl ether (CH$_3$OCH$_3$) reacting with H$^+$.
\cite{vazart2020} has reviewed all these reactions from the theoretical and experimental point of view and concluded that only the first two routes are viable (i.e. involving CH$_3$CH$_2$ and CH$_3$CH$_2$OH, respectively), while the last two (i.e. those involving CH$_3$OH and CH$_3$CHOH$^+$) are inefficient in the ISM conditions.
Finally, the observed correlation between the derived abundances of acetaldehyde and ethanol in warm and hot sources is in favor of acetaldehyde being the ethanol daughter \citep{vazart2020}.

For the grain-surface formation routes, the situation is more debated.
Since the end of last millennium, experiments have found that acetaldehyde is formed in UV-illuminated ices consisting of water, CO, methanol and methane \citep[e.g.][]{Hudson1997,Bennett2005,Oberg2010,MartinDomenech2020} or other components \citep[e.g.][]{Chuang2021}.
Triggered by these experiments, \cite{Garrod2006} proposed the formation of acetaldehyde from the combination of the radicals HCO and CH$_3$ on the dust grain icy surfaces, when the grain temperature increases enough to make the two radicals mobile. 
However, a first study by \cite{Enrique-Romero2016} suggested that the reaction of the two radicals on an amorphous water ice surface actually does not end up into the formation of acetaldehyde but rather into carbon monoxide (CO ) and methane (CH$_4$).
Subsequent and more accurate studies confirmed that the acetaldehyde formation on amorphous water surfaces may or may not occur and is always in competition with the CO + CH$_4$ formation \citep[][]{Rimola2018,Enrique-Romero2019,Enrique-Romero2020}.
Similar conclusions were reached when considering the reaction on CO-rich ices \citep{Lamberts2019}.
Finally, the most recent theoretical study by \cite{Enrique-Romero2021} found that the efficiency of the formaldehyde formation by the coupling of HCO and CH$_3$ is a strong function of the grain temperature and the diffusion energies of the two radicals, with the possibility to being practically zero if their mobility is relatively large (i.e., a diffusion/BE ratio less than 0.3).
In agreement with these theoretical predictions, a very recent and sophisticated study by \cite{Gutierrez-Quintanilla2021} found no formation of acetaldehyde on an amorphous water ice enriched with HCO and CH$_3$.
In their experiment, \cite{Gutierrez-Quintanilla2021} trapped radicals, formed via photolysis of methanol ice, in an argon matrix at 14 K and identified them via Electronic Paramagnetic Resonance (EPR) spectroscopy. 
Once liberated from the matrix, the various radicals combined giving rise to several species, but not acetaldehyde.
More recently,  non energetic processes on the grain surfaces have been found to possibly produce acetaldehyde in CO-rich ices \citep[e.g.][]{,Fedoseev2022}.
Adding confusion to this situation, \cite{Hudson2020} noticed that the IR bands, on which the identification (frequencies) and quantification (band strengths) of acetaldehyde in the various experiments relied, were often incorrect.
Finally, this work also emphasized the almost impossibility for astronomical observations to be able to identify frozen acetaldehyde.

Whatever the formation mechanism, either in the gas-phase or on the grain surfaces, acetaldehyde would freeze onto the cold dust grain mantles and would be removed from them only when the dust temperature reaches the acetaldehyde sublimation temperature, which is governed by its binding energy (BE: sometimes it is also called desorption energy) to the water ice, and its pre-exponential factor (see \S ~\ref{sec:discussion}) .
Please note that the sublimation temperature can be reached due to both thermal and non-thermal processes, such as the desorption caused by the interaction of the cosmic-rays, which permeate the Milky Way, with the dust grains. 
In the latter case, for example, only a fraction of the grain is heated up and acetaldehyde would be desorbed if the reached temperature is larger than its sublimation temperature.
Therefore, in practice, whether acetaldehyde is in the gas-phase (where it can be observed via its rotational lines) or frozen onto the grain mantles (where it has not been detected so far) is completely governed by its BE.
Moreover, BE always enters in an exponential form, so that its accurate estimation is essential to properly assess whether and how much acetaldehyde would be gaseous.

So far, there are very few specific studies on the acetaldehyde BE.
To our best knowledge, \cite{Wakelam2017BE} is the first theoretical study, but it only considered one water molecule to simulate the interstellar water ice.
These authors found a BE equal to $\sim$5400 K.
Recently, \cite{Corazzi2021} reported an experimental study of the BE on olivine surfaces covered by an ice of pure acetaldehyde or mixed with water.
They found that the acetaldehyde BE on a mixture of iced water-acetaldehyde is 3079 K, substantially different from the Wakelam et al.'s estimate. In a very recent combined theoretical and experimental study, \cite{Molpeceres2022} obtained, for the desorption of acetaldehyde on non-porous amorphous solid water surfaces, the values of 3624 and 3774 K, respectively. These values are relatively similar to the Corazzi et al's value, but still far from that of Wakelam et al.'s one. There is, therefore, a need to clarify the origin of this difference as well as to improve both experimental and theoretical estimates.

In this work, we combine state-of-the-art experimental (temperature programmed desorption, TPD) and theoretical (quantum chemical computations) techniques with the aim to provide possibly the most accurate estimate of the acetaldehyde BE on water ice, simulating at best the ISM conditions.
The article is organised as follows.
Sec. \ref{sec:experiments} reports the experiments, Sec. \ref{sec:calculations} the theoretical computations and Sec. \ref{sec:discussion} discusses the results and comparison of the two methods, as well as the astrophysical implications.

\section{Laboratory experiments}\label{sec:experiments}
The experiments were performed with the FORMOLISM (FORmation of MOLecules in the InterStellar Medium) set-up. The apparatus is meant to study the reactivity of atoms and molecules on surfaces of astrophysical interest, under the conditions of temperatures and pressures similar to those in the ISM. The practicalities of the experimental setup are described here but more details are given in previous works \citep{amiaud2007, Congiu2012}.

\subsection{Methods}
FORMOLISM is composed of an ultra-high vacuum (UHV) stainless steel chamber with a pressure of a few 10$^{-11}$ mbar. The sample holder is located in the center of the main chamber and it is thermally connected to a closed cycle Helium cryostat. The temperature of the sample is measured in the range of 10--800 K by a calibrated titanium diode connected to the sample holder. This one is made of a 1 cm diameter copper block (18 mm long), which is covered with a highly oriented pyrolytic graphite (HOPG, ZYA-grade) sample. The HOPG is a model of an ordered carbonaceous material, carbon atoms in a hexagonal lattice, mimicking some aspects of interstellar dust grains analogues. The HOPG grade (10 mm diameter-2 mm thickness) was firstly dried in an oven at about 100°C for two hours, and then cleaved several times in air using the "Scotch tape" method at room temperature to yield several limited defects and step edges \citep{chaabouni2020}.
The HOPG was glued directly onto the copper finger and dried in ambient conditions for about an hour before inserting it in the chamber and closing the set up. Once high vacuum is reached (<10$^{-7}$ mbar) in  few minutes the sample is dried under vacuum during 3 hours at 350 K. Then the HOPG sample and the whole system is baked out at 100°C for few days in order to remove any adsorbed contaminants and reach the UHV base pressure. Before starting any experiment, the system is slowly heated up to 720 K to maximise the degassing rate and ensure a very clean sample surface.

FORMOLISM is equipped with a quadrupole mass spectrometer (QMS). It allows both the detection of species desorbing from the sample during a TPD (temperature programmed desorption) and, when placed in front of the beam-line, the characterization and the calibration of molecular beams.

The CH$_3$CHO (Sigma Aldrich, purity higher than 99.5 \%) molecular beam is prepared in a triply differentially pumped beam-line aimed at the sample holder. It is composed of three vacuum chambers connected together by tight diaphragms of 3 mm diameters. Molecular pressure injected in the gas line is of about 1.4 mbar. In the expansion chamber (first stage) the pressure is in the 10$^{-5}$ mbar range, while in the main chamber the rise of the pressure is not measurable (<10$^{-11}$ mbar).

The CH$_3$CHO on HOPG experiments have been used as reference and preliminary studies before investigating the desorbing behaviour of the molecule on two different D$_2$O ice substrates formed on the sample. In this study, we have deposited CH$_3$CHO molecules on HOPG by using only one beam-line oriented at 57° relatively to the surface of the sample, while D$_2$O ice films have been deposited on HOPG by using a separated channel and a micro-capillary array doser moved close to the sample holder. It allows to control the ice deposition rates and to avoid to raise the pressure in the whole chamber. The two different D$_2$O ices are, respectively, compact non-porous amorphous deuterated water (ASW-c) and (poly)crystalline deuterated ice (PCI). ASW-c is formed when the sample is held at 110 K, while PCI layers have been formed by depositing D$_2$O at 110 K, heating until 150 K with a ramp rate of 5 K/min and waiting for few minutes to be sure that the crystallization has happened, as detected by the change in the desorption rate \citep{Speedy1996}. 
For both ices, about 30 monolayers (ML)  have been deposited on HOPG while the pressure in the main chamber never exceeded $5 \times 10^{-8}$ mbar. 
After the CH$_3$CHO deposition phase at 45 K, we used the TPD technique by warming-up the sample and detecting the desorbing species from the surface through mass spectrometry. 
TPD were obtained from 45 to 220 K for the CH$_3$CHO on HOPG experiments and only to 140 K for the CH$_3$CHO on water ice ones, in order to avoid D$_2$O desorption after 150 K. 
In all cases, a linear heating rate of 12 K/min has been programmed to heat up the sample. 
Because of the poor thermal contact between the sample holder and the copper block, the temperatures registered during the TPD were re-calibrated by depositing different gases of known BEs (O$_2$, Kr, D$_2$O). 
The accuracy is about 1 K for the values of temperature below 125 K and 0.2 K for the ones above. The actual new ramps obtained after calibration are taken into account in the simulations and analyses.

With the QMS, the same molecule can be detected in different m/z corresponding to the fragmentation of the parent molecules after the electron impact (here at 30 eV). The main fragments under study here are m/z=44 (molecular mass of CH$_3$CHO$^+$) and  m/z=29 (HCO$^+$, the main fragment of acetaldehyde). 
The desorption and degree of crystallization of D$_2$O ice films have been followed through the signal of  m/z=20.

\subsection{Results}\label{subsec:exp-results}
\begin{figure}
   \centering
   \resizebox{\hsize}{!}{\includegraphics{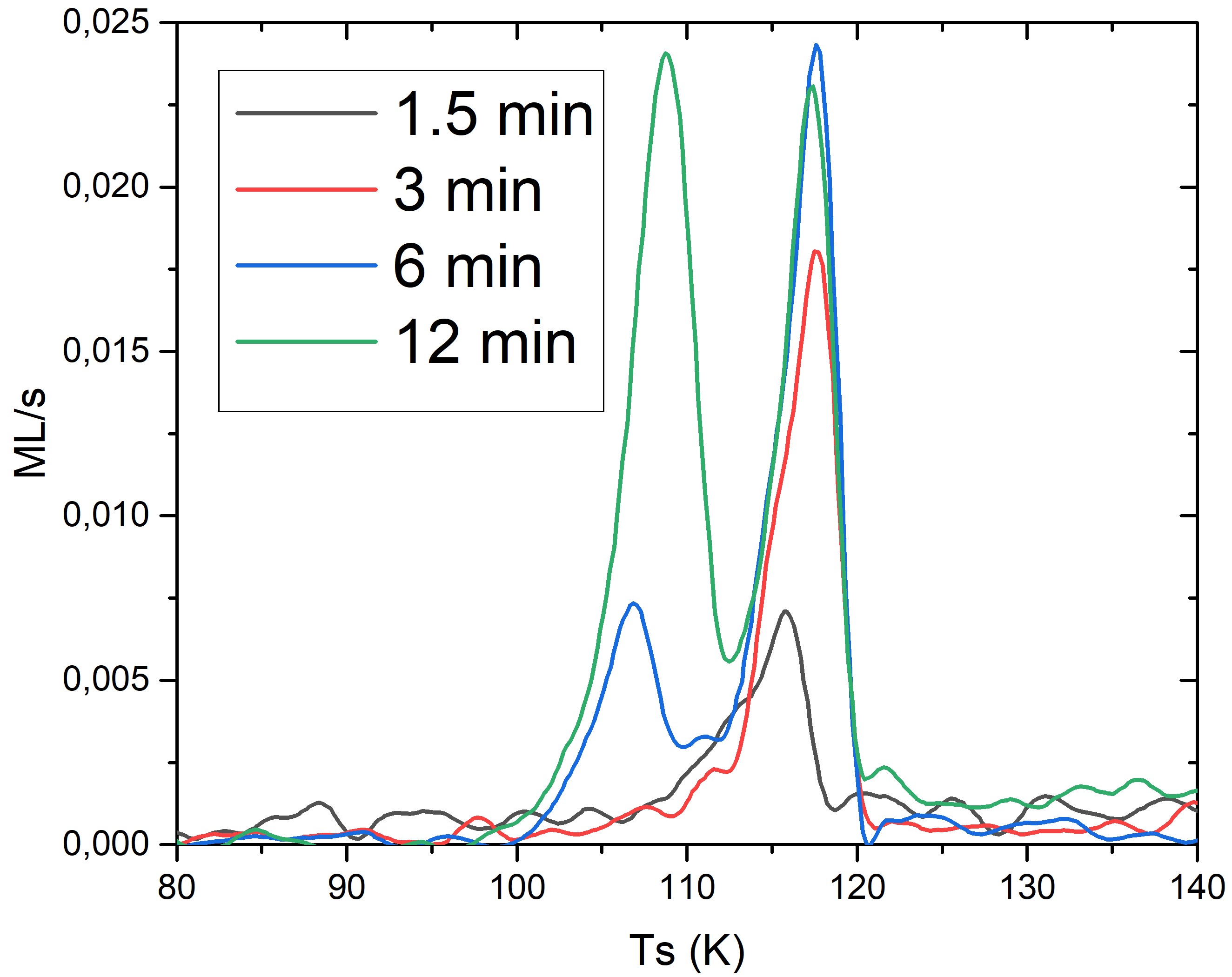}}
   \caption{Set of four smoothed TPDs (m/z=44) for different time depositions of acetaldehyde on HOPG: 1.5 min (0.4 ML), 3 min (0.8 ML), 6 min (1.7 ML), 12 min (3.3 ML).}
    \label{fig:Famille TPD CH3CHO SUR HOPG}
    \end{figure}

Experiments on HOPG have been used mainly as reference, in order to understand the behavior of the acetaldehyde adsorption and the deposition time needed to have a monolayer on the substrate. 
Seven measurements corresponding to increasing deposition times (1, 1.5, 3, 4, 5, 6, 12 minutes) have been carried out on HOPG to study the evolution of the TPDs from a sub-monolayer regime to a multilayer one. 
Fig. \ref{fig:Famille TPD CH3CHO SUR HOPG} shows only four TPD curves belonging to four different deposition times using the same first stage pressure to introduce a constant flux of  CH$_3$CHO in the system. 
There are two main features appearing in the graph: a low temperature peak between 105-110 K, and a second peak  centered around 117.5 K. 
At low deposition times, only the peak at higher temperature appears (black and red curves corresponding to 1.5 and 3 minutes). 
It initially increases in intensity with coverage but saturate for longer depositions. 
After 6 minutes of deposition, the low temperature peak clearly appears. 
The TPD after 12 min deposition shows both peaks equally intense, although the low temperature peak is slightly broader. 
The high temperature peak is attributed to sub-monolayer regime and is first populated, and, when the monolayer is saturated and the multilayer regime starts, the second peak at lower temperature appears. 
By following the TPD experiments on HOPG, a monolayer of molecules on the substrate is believed to form after ca. 4 minutes of deposition. 
We define this time to calibrate the rest of the analyses. 
As HOPG and ASW-c have about the same surface density of sites, the same time deposition is used to characterise the coverage on ice, and to derive the BEs of CH$_3$CHO as a function of the coverage.

\begin{figure}
   \centering
   \resizebox{\hsize}{!}{\includegraphics{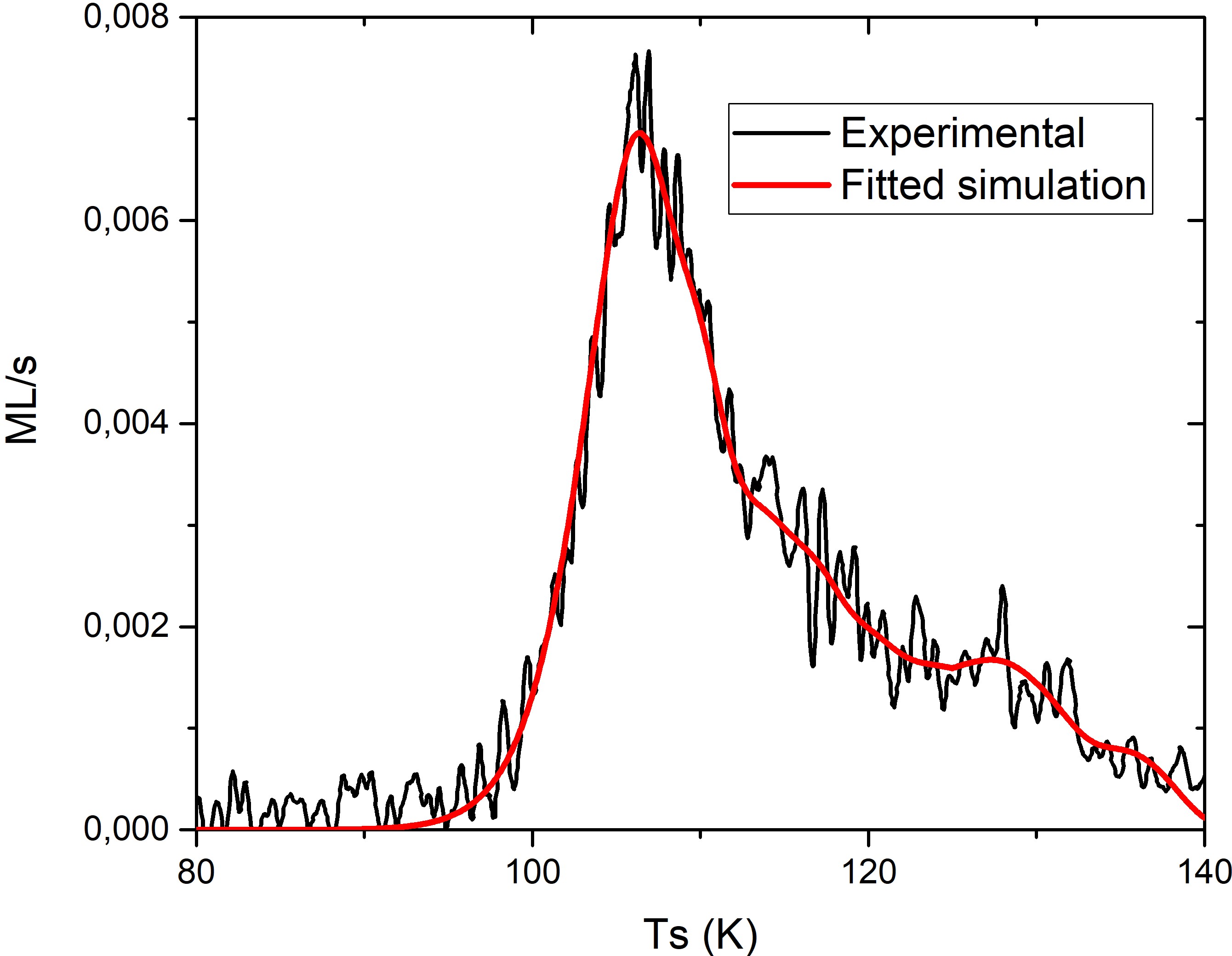}}
   \caption{Simulation of eleven TPDs (red line) used to fit the experimental TPD (black line, raw data not smoothed) obtained from 4 min (1 ML) of CH$_3$CHO  deposition on PCI  (pre-exponential factor=$1.1 \times 10^{18}$ s$^{-1}$, E$_{min}$=4500 K, E$_{max}$=6000 K by steps of 150 K).}
    \label{fig:Famille TPD CH3CHO SUR PCI}
    \end{figure}

In order to derive the BEs on HOPG and on the two different ices, the three TPDs for the 4 min depositions on HOPG, ASW-c and PCI have been selected. 
By proceeding this way, the interaction between a monolayer of deposited CH$_3$CHO and each substrate can be studied. 
Fig. \ref{fig:Famille TPD CH3CHO SUR PCI} shows the results for one of them, i.e., the experiment of 4 min deposition of CH$_3$CHO on PCI, and the simulation obtained by fitting the data. 
The procedure is detailed in \cite{Chaabouni2018}. 
The calculation uses eleven independent TPDs, starting from 4500 K, with an energy gap between each TPD of 150 K (i.e. 4500, 4650, 4800 K...). 
Each individual TPD is calculated using simple Arrhenius kinetics, named the Polanyi-Wigner equation:  
 
\begin{equation}
r(T) =-\frac{dN}{dT}=\ A N^n \exp{\left(-\frac{E_{des}}{T} \right)}
\label{eq:eq1}
\end{equation}

where $r(T)$ is the desorption rate (in ML s$^{-1}$), $N$ is the number density of molecules adsorbed on the surface (ML), and $n$ is the order of the desorption equal to 1 in our case. 
$A$ is the pre-exponential factor (s$^{-1}$), $T$ is the  temperature of the surface (in K), and $E_{des}$ is the activation energy for desorption (in K).
If there is no reorganisation of the surface\footnote{This is 100\% true on graphite, but this is an assumption/approximation in case of acetaldehyde on water.}, $E_{des}$ is equal to BE.
First order desorption TPD profiles are independent of the surface population, so that each independent BE can be weighted, and finally a distribution of BEs can be found. 
We have set the pre-exponential factor at $1.1 \times 10^{18}$ s$^{-1}$ (see \S ~\ref{sec:discussion}).
The arbitrary choice of binning the energy distribution (basis of eleven BEs) have been optimized with the aim to have a good fitting of all the three experiments under analysis. 
Fig. \ref{fig:Famille TPD CH3CHO SUR PCI} gives a qualitative idea of the simulation for the case of the PCI substrate. 
The best values found for the eleven TPDs, also known as energy distribution set for each substrate, are reported in Table \ref{tab:my-table}. 
The values are here expressed as a function of the population, in percentage, having that specific BE. 
We estimate the accuracy of the method to be around a few percents, mostly due to the noise-to-signal ratio and the presence of a small background noise.

\begin{table}
\caption{Desorption energy values of the eleven simulated TPDs obtained for three sets of 4 min CH$_3$CHO depositions: on HOPG, ASW-c and PCI (pre-exponential factor:$1.1 \times 10^{18}$ s$^{-1}$).}
\label{tab:my-table}
\resizebox{\hsize}{!}{%
\begin{tabular}{l|l|l|l|}
\cline{2-4}
 & HOPG & c-ASW & PCI \\ \hline
\multicolumn{1}{|l|}{E$_{des}$ (K)} & Population (\%) & Population (\%) & Population (\%) \\ \hline
\multicolumn{1}{|l|}{4500} & 0 & 0 & < 4 \\ \hline
\multicolumn{1}{|l|}{4650} & 10 & 16 & 28 \\ \hline
\multicolumn{1}{|l|}{4800} & 0 & 32 & 27 \\ \hline
\multicolumn{1}{|l|}{4950} & 0 & 10 & 12 \\ \hline
\multicolumn{1}{|l|}{5100} & 34 & 12 & 12 \\ \hline
\multicolumn{1}{|l|}{5250} & 66 & 7 & 9 \\ \hline
\multicolumn{1}{|l|}{5400} & 0 & 8 & 6 \\ \hline
\multicolumn{1}{|l|}{5550} & 0 & 9 & 6 \\ \hline
\multicolumn{1}{|l|}{5700} & < 4 & 6 & 8 \\ \hline
\multicolumn{1}{|l|}{5850} & < 4 & 7 & < 4 \\ \hline
\multicolumn{1}{|l|}{6000} & < 4 & < 4 & < 4 \\ \hline
\end{tabular}%
}
\end{table}
The three energy distributions are displayed in Fig. \ref{fig:Energy distribution of 4 min TPD for HOPG, ASW and PCI}. 
We omit the populations below 4 percents. 
One can observe the main differences because of the different substrate. 
For HOPG, there is not an actual distribution, but it rather shows almost only two main energy values, which are associated with the sub-monolayer regime (at 5100--5250 K). 
As can be noticed by the 10\% population at 4650 K, for the 4 min depositions the multilayer regime slightly started already. 
This is in agreement with the accuracy of the determination of the flux, which is usually estimated around 15$\%$.

The distributions of BEs of CH$_3$CHO on ices present different characteristics, signature of the more disordered nature of water ice (even for the PCI case). 
There is a larger distribution of population covering almost all the possible energy values proposed by the simulation. The ASW-c substrate presents the largest distribution of BEs.  
The energy distribution histogram associated with the desorption of CH$_3$CHO on PCI presents, instead, a more peaked distribution towards lower values of energy (mainly below 4900 K). 
This is due to the reduced number of combination of water molecules on the surface. The disorder here is provoked by the possibility of having an O or an H in the vicinity of the adsorption sites, even if the collective structure is cubic. On the contrary, ASW-c have a more disordered nature and, accordingly, the adsorption sites may have a more variable number of water molecules strongly interacting with CH$_3$CHO (see below).
Finally, we calculate the weighted average for each case to estimate the difference in BEs for the different substrates. We obtain the following result: 5150 K for HOPG (range: 4650--5250 K), 5080 K for ASW-c (range: 4650--5850 K) and 4990 K for PCI (range: 4650--5700K).
This corresponds to what one can expect by looking at the trend in the desorption profiles.
\begin{figure}
   \centering
   \resizebox{\hsize}{!}{\includegraphics{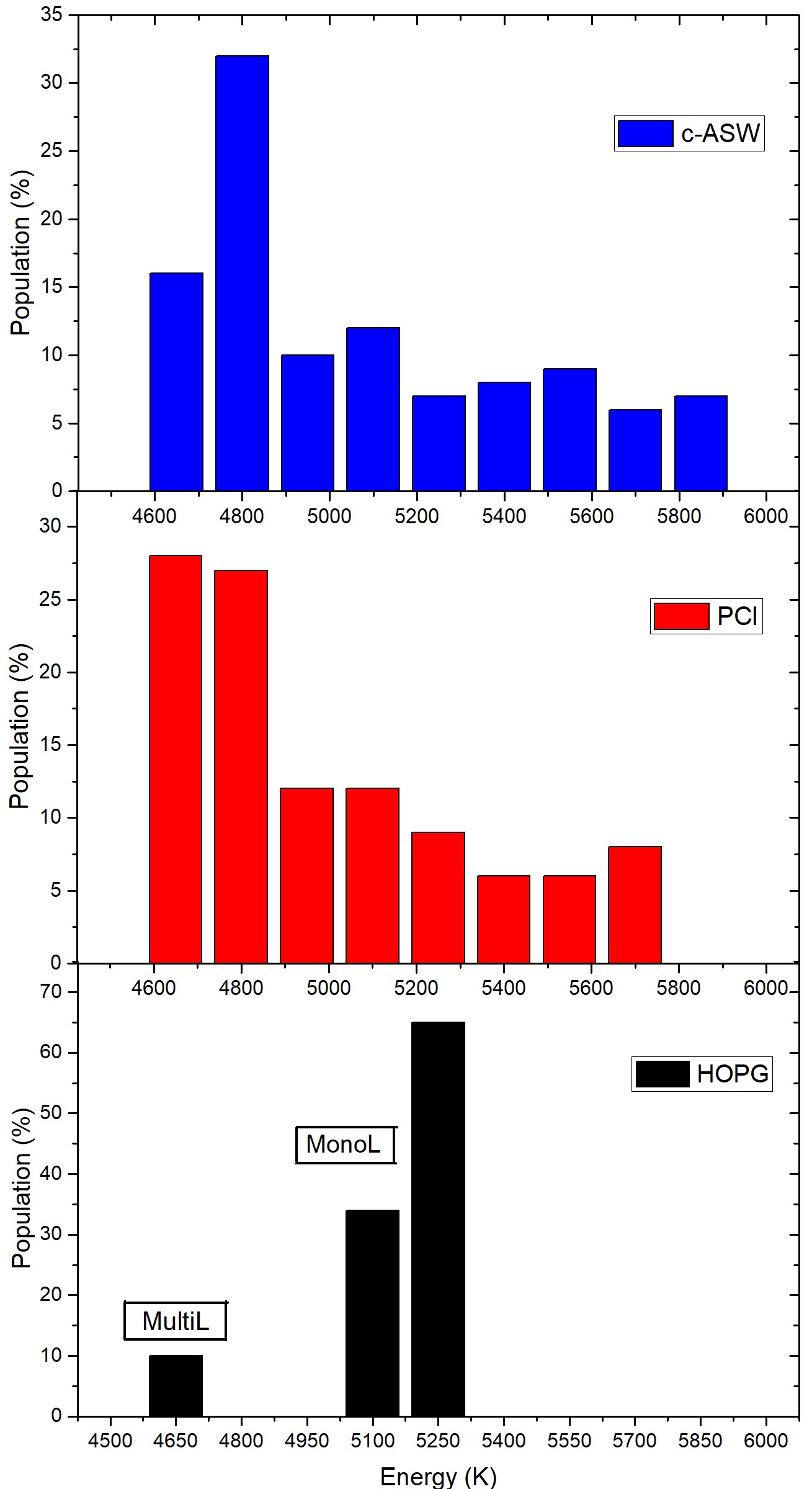}}
   \caption{Energy distributions of eleven simulated TPDs for the 4 min CH$_3$CHO depositions on HOPG, ASW-c and PCI taken from Table \ref{tab:my-table}.}
    \label{fig:Energy distribution of 4 min TPD for HOPG, ASW and PCI}
    \end{figure}

\section{Quantum chemistry calculations}\label{sec:calculations}
Simulations of the adsorption of an acetaldehyde molecule on crystalline and amorphous ice models, as well as a conmponent of a pure acetaldehyde surface have been carried out in order to i) gain insights of the adsorption process at an atomic level, and ii) calculate BEs of the species on ices with the purpose to make comparisons with TPD experiments.       
\subsection{Methods}
\subsubsection{Computational details}
All the calculations were performed with the ab initio CRYSTAL17 code \citep{dovesi2018quantum}. The code can simulate any kind of system, from non periodic molecules to full periodic 3D crystalline systems, employing atom centered Gaussian basis sets for the description of the electronic structure. In this work, we carried out DFT-based static optimizations using the BFGS optimization algorithm, relaxing both the cell parameters and the atomic positions. These calculations were performed with the PBEsol0-3c functional \citep{PBE0sol-3c}, which makes use of an Ahlrichs' polarised valence-single zeta basis set in order to reduce the computational cost. Then, on the PBEsol-3c optimized geometries, single point energy calculations were carried out with the B3LYP functional \citep{Becke1993, LYP88} added with the Grimme's D3(BJ) correction (to account for dispersive forces \citep{Grimme2010, Grimme2011}) with the aim to refine the BEs at a higher level of theory. These refinement calculations have been performed using an Ahlrichs' VTZ basis set added with a double set of polarization functions. For the sake of accuracy, the calculations carried out with the B3LYP-D3(BJ) method have been corrected for the basis set superposition error (BSSE), adopting the \textit{a posteriori} counterpoise correction.   

Vibrational harmonic frequencies were calculated at the PBEsol0-3c level using a finite differences method. A partial Hessian approach was used to reduce the computational cost of the calculations. Thus, the vibrational frequencies were calculated on the optimized geometries only for a fragment of the entire system, which included the acetaldehyde molecule and the closest water molecules interacting with it. From this vibrational frequency calculations, the zero point energy (ZPE) correction terms were included in the calculations of the BEs. 

\subsubsection{Water ice and pure acetaldehyde periodic structures}
\begin{figure*}
   \centering
   \resizebox{\hsize}{!}{\includegraphics[ width=\textwidth]{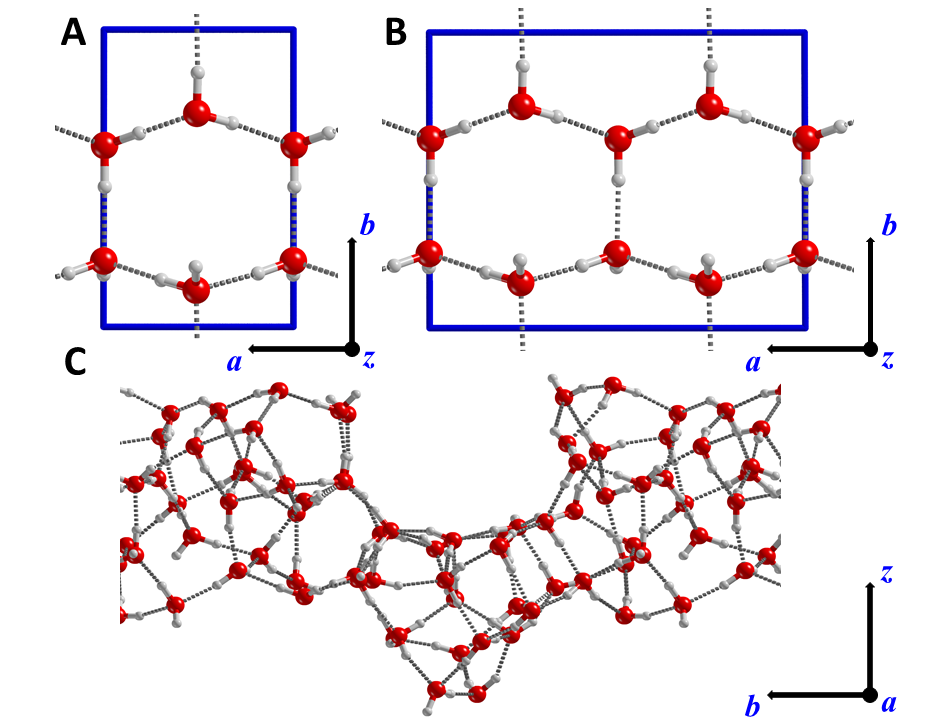}}
   \caption{\textit{Panel A:} Top view of the 1x1 unit cell of the P-ice (010) surface. 
   \textit{Panel B:} Top view of the 2x1 supercell of the P-ice (010) surface.  
   \textit{Panel C:} Side view of the amorphous ice surface.
   The employed 1x1 cell and 2x1 supercell of panels A and B are highlighted in blue.}
    \label{fig:ice_surfaces}
    \end{figure*}

The crystalline and amorphous water ice surface models used in this article have already been described in \cite{Ferrero2020} and are reported in Fig. \ref{fig:ice_surfaces}. Briefly, the crystalline slab model derives from the (010) surface of the proton-ordered P-ice. For this surface, two different unit cells have been used, the 1x1 and the 2x1 (Fig. \ref{fig:ice_surfaces}A and B, respectively), this way covering both high and low coverage regimes. 
The amorphous water ice surface was constructed by joining different water clusters from \cite{Shimonishi_2018}, in which, upon imposing periodicity and optimising the structure, the resulting surface model presents a cavity (Fig. \ref{fig:ice_surfaces}C). 

Since the crystalline structure of pure acetaldehyde is not available yet, in order to simulate a pure acetaldehyde slab we resorted to the crystal structure prediction algorithm. First, the acetaldehyde molecule was optimized in gas phase 
with the B3LYP-D3(BJ) method combined with the optimized def2-SVP basis set with the Gaussian 09 \citep{gaussian0920091} program. This gas-phase optimized structure was used as input parameter to derive the pure acetaldehyde periodic solid state structures, which was achieved by using the programs of DFTB$+$ (version 20.2.1) \citep{hourahine2020dftb+} including Self Consistent Charge \citep{elstner1998self} and the D4 dispersion model by Grimme \citep{caldeweyher2019generally, caldeweyher2017extension}, and VASP 5.4.4 \citep{kresse1993ab, kresse1994ab, kresse1996efficiency, kresse1996efficient} using PAW PBE pseudopotentials \citep{kresse1999ultrasoft}.

Solid-state structure prediction was performed using the USPEX (version 10.4.1) evolutionary algorithm  \citep{lyakhov2013new, oganov2006crystal, oganov2011evolutionary}, with embedded the topology crystal structure generator \citep{bushlanov2019topology}, employing the DFTB$+$ code for geometry optimizations. The first generation of crystal structures were randomly generated, while subsequent generations were obtained by applying the genetic algorithm embedded in USPEX coupled with different variation operators, in particular rot-mutation and lattice mutation. The search space was limited from 1 to 4 molecules per cell. Calculations proceeded for 23 generations and 570 structures, in which USPEX identified the best structure, which contains 2 molecules per cell and a final volume of 118.77 \AA $^3$. Since periodic DFTB$+$ calculations tend to excessively shrink the unit cell, subsequent calculations with USPEX but this time employing VASP calculations were performed using as initial seeds the best structures found by DFTB$+$. This proceeded for 28 generations and 723 structures, resulting in the final optimized structure characterized by a $=$ 4.960 \AA, b $=$ 5.655 \AA, c = 4.536 \AA, and $\alpha$ $=$ 89.55$^{\circ}$, $\beta$ $=$ 89.70$^{\circ}$, $\gamma$ $=$ 127.17$^{\circ}$ (with a final volume of 127.17 \AA $^3$)

\begin{figure}
   \centering
   \resizebox{\hsize}{!}{\includegraphics{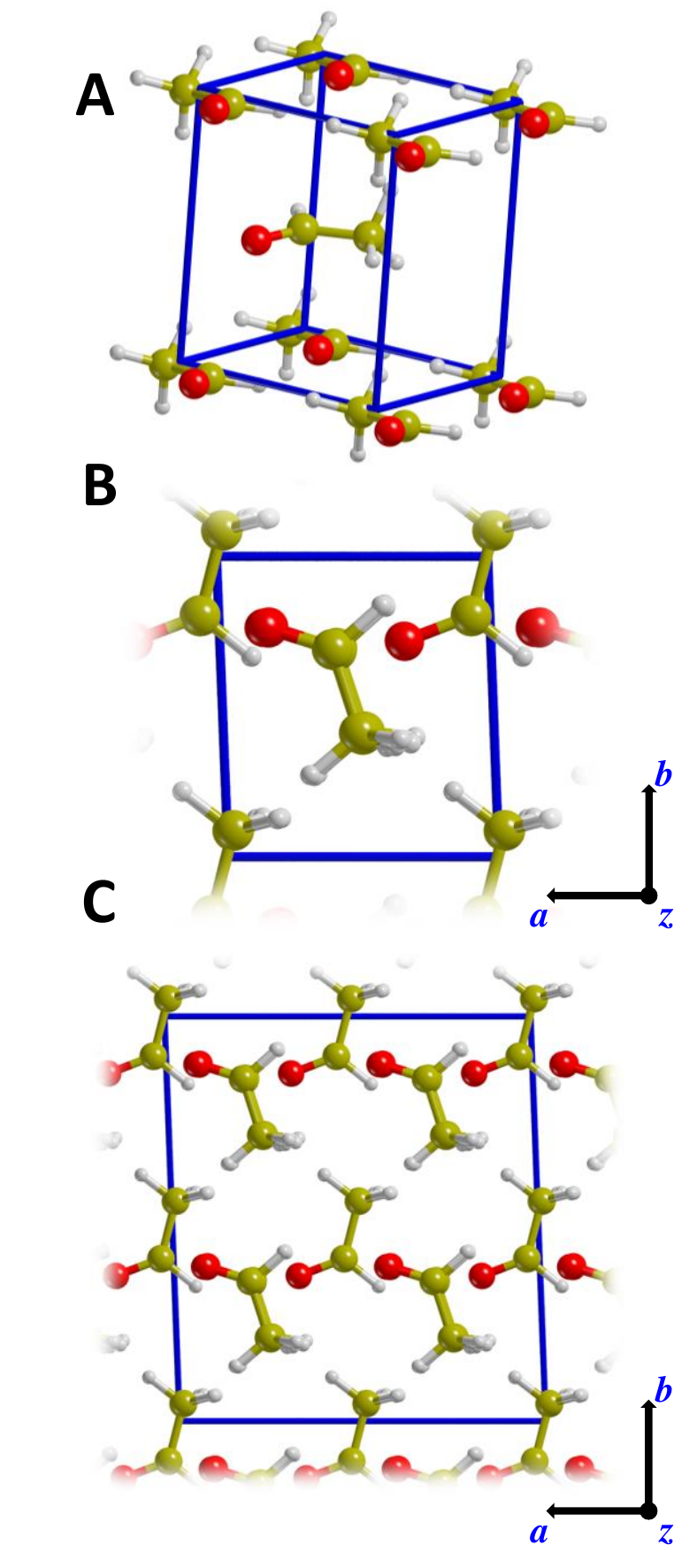}}
   \caption{\textit{Panel A:} PBEsol0-3c optimized structure of the bulk model for pure acetaldehyde. 
   \textit{Panel B:} Top view of the 1x1 unit cell of the (010) pure acetaldehyde slab surface model. 
   \textit{Panel C:} Top view of the 2x2 supercell of the (010) pure acetaldehyde slab surface model. 
   The employed unit cells are highlighted in blue.}
    \label{fig:acetaldeide}
    \end{figure}

The resulting bulk structure has been optimized at the PBEsol0-3c level and it is depicted in Fig. \ref{fig:acetaldeide}A. Once we obtained the bulk, we cut it along the (010) direction to obtain a 2D periodic slab model, which does not posses a large dipole along the non periodic direction (\textbf{z} axis (Fig. \ref{fig:acetaldeide}B and C).

\subsection{Results}\label{subsec:comp-results}
\subsubsection{Adsorption of acetaldehyde on ice structures}
\begin{figure}
   \centering
   \resizebox{\hsize}{!}{\includegraphics{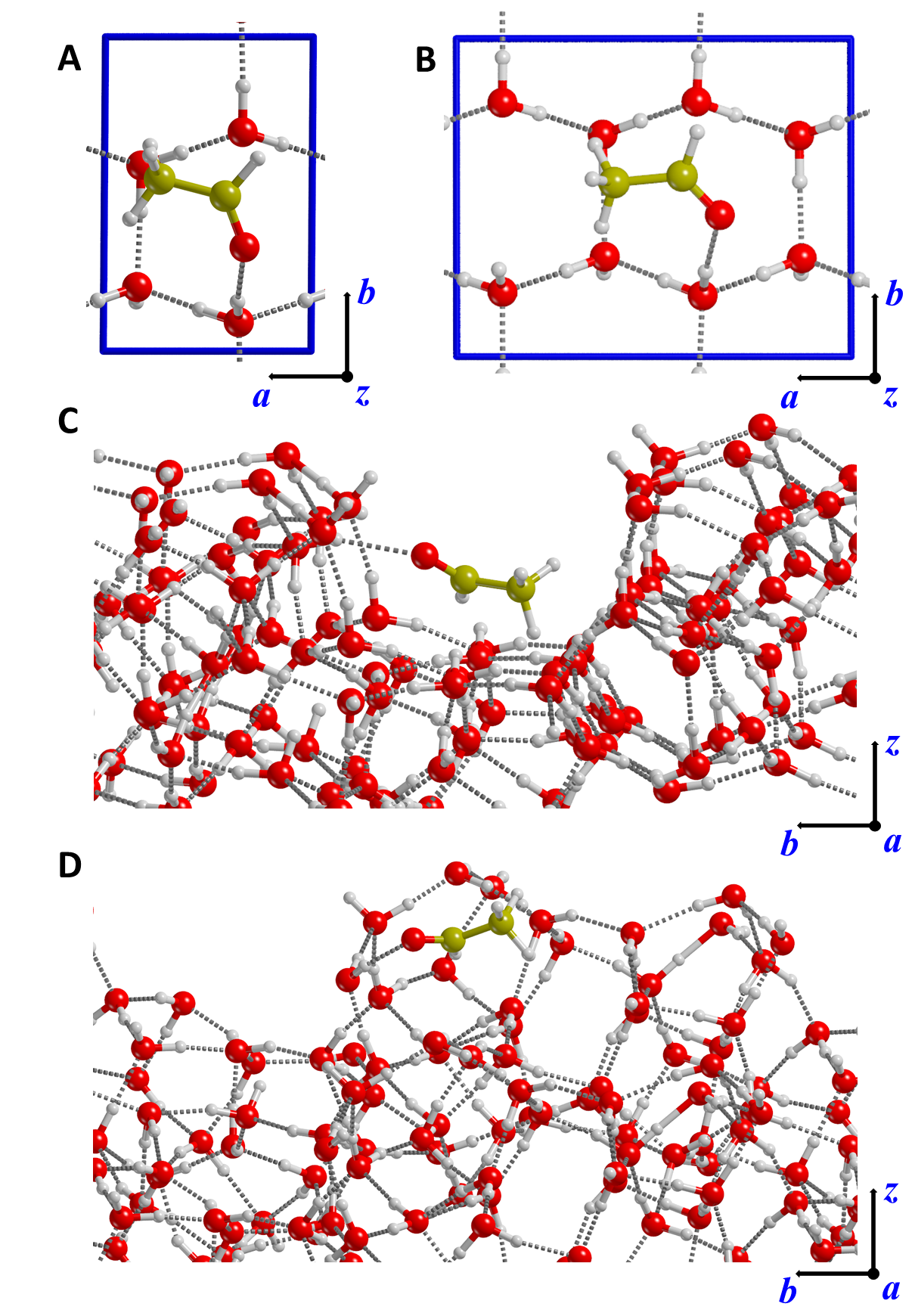}}
   \caption{\textit{Panel A:} Top view of the optimized geometry of one acetaldehyde molecule adsorbed on the 1x1 cell of the P-ice (010) surface model. 
   \textit{Panel B:} Top view of the optimized geometry of one acetaldehyde molecule adsorbed on the 2x1 supercell of the P-ice (010) surface model. 
   The 1x1 cell and the 2x1 supercell are highlighted in blue. 
   Side view of the optimized geometries of acetaldehyde adsorbed on the amorphous ice surface model presenting the highest (\textit{Panel C}) and the lowest (\textit{Panel D}) BEs.}
    \label{fig:adsorption}
    \end{figure}

We first simulate the acetaldehyde adsorption process by placing the molecule on the 1x1 crystalline surface, in which a single hydrogen bond (H-bond) between a dangling hydrogen (dH) of the ice surface and the O atom of acetaldehyde is established. The size of the 1x1 unit cell is very small (it only contains one dH) and accordingly  adsorption of acetaldehyde ends up with a structure resembling a mono-layer coverage adsorption situation (see panel A of Fig. \ref{fig:adsorption}), in which lateral interactions between acetaldehydes of adjacent unit cells take place due to the periodic boundary conditions. By using the 2x1 unit cell, such lateral interactions are removed, giving rise to a low coverage regime, in which a single acetaldehyde is adsorbed on the surface (see panel B of Fig. \ref{fig:adsorption}). 
The obtained ZPE-corrected binding energies, BE(0), are 7253 K and 5194 K for the 1x1 and 2x1 unit cell cases, respectively (reported in Tab. \ref{tab:BEs}). 
Since the amorphous ice is larger (60 water molecules) and presents a disordered surface morphology, this surface model presents more adsorption sites. We sampled some of them by placing acetaldehyde molecule in different starting positions and then optimising the structures. Out of a total of 14 optimizations, we obtained 9 different minima, in which in all the cases acetaldehyde establishes H-bond interactions with dH of the amorphous surface. The different calculated BE(0) values are reported in Table \ref{tab:BEs}, which spans the ca. 2800--6000 K range. Panels C and D of Fig. \ref{fig:adsorption} show the optimized structures for the complexes presenting the highest and the lowest BE(0) values, respectively.

\begin{table}
\caption{Calculated (ZPE corrected) BE(0) (in K) of acetaldehyde on the crystalline and the amorphous ice surface models.}
\label{tab:BEs}
\begin{tabular}{cc}
\hline
\multicolumn{1}{l}{\textbf{Crystalline ice}} & \textbf{BE(0) } \\ \hline
\multicolumn{1}{l}{1x1 cell} &  7253  \\ 
\multicolumn{1}{l}{2x1 cell} & 5194 \\ 
\hline
\multicolumn{1}{l} {\textbf{Amorphous ice}}  &   \\ 
\hline
\multicolumn{1}{l}{Site$_1$} & 5208  \\
\multicolumn{1}{l}{Site$_2$} & 3366  \\ 
\multicolumn{1}{l}{Site$_3$} & 5700 \\ 
\multicolumn{1}{l}{Site$_4$} & 4708  \\ 
\multicolumn{1}{l}{Site$_5$} & 2809  \\ 
\multicolumn{1}{l}{Site$_6$} & 4028  \\ 
\multicolumn{1}{l}{Site$_7$} & 3450  \\
\multicolumn{1}{l}{Site$_8$} & 4597  \\
\multicolumn{1}{l}{Site$_9$} & 6038  \\ \hline
\end{tabular}
\end{table}

\subsubsection{Adsorption of acetaldehyde on pure acetaldehyde surface}
A single acetaldehyde molecule was adsorbed on the 2x2 super cell of the (010) pure acetaldehyde slab model. Its binding energy BE(0) was computed similarly as described for the adsorption on the water ice structures. The calculated BE(0) for this case is 2650K. This value is a lower limit of the BE(0), as the isolated molecule is engaged in very few interactions with the companion surface. The upper limit of the BE is the cost needed to extract a single acetaldeyde molecule from the external layer of the (010) surface. In this case, the  BE(0) is a compromise between the rupture of a consistent number of intermolecular interactions holding acetaldehyde  within the slab and the geometry relaxation of the cavity created in the surface. The binding energy BE(0) was, therefore, calculated by subtracting the energy of the relaxed surface with the cavity \emph{plus} the energy of the extracted acetaldehyde from the energy of the pristine perfect surface. The result is a BE(0)=5692 K, that is, almost twice as large as the acetaldehyde BE(0) at the clean (010) surface (\emph{vide supra}). 


\section{Discussion}\label{sec:discussion}

\subsection{The importance of the pre-exponential factor}\label{subsec:preexp-factor} 

Before discussing the chemical physics, or any methodological bias, it is important to be able to compare the values derived by the experimental and theoretical methods, which in the case of desorption is not straightforward. 
To understand this point, we will first recall the framework in which the BE values are obtained in each approach. 
Unfortunately, the three communities (laboratory, quantum chemistry, and astrochemistry) have their own underlying definitions, which we will discuss first. 

\noindent
\textit{The experimental point of view:} 
Experiments do not directly measure a desorption energy, but a desorbing flux, and, by applying Eq. \ref{eq:eq1} (see \S ~\ref{sec:experiments}), a BE is derived (which is supposed to be equal to the desorption energy in most of the cases). 
However, to do so, one needs to set a value for the pre-exponential factor, since the equation has two parameters: the pre-exponential factor A and the binding energy BE (if the desorption order is set to $n$=1). 
Here, following the recommendation by \cite{Minissale2022}, we adopt the derivation of the pre-exponential factor based on the Transition State Theory (TST). 
This theory takes into account both the rotational and translation partition functions of the desorbing molecules and, therefore, allows including in the calculation of the pre-exponential factor, $\rm {A_{TST}}$, the entropic effect associated with the kinetic desorption rate. 
We briefly describe the salient points of the calculation of $\rm{A_{TST}}$ \citep[more details can be found in][]{tait2005b}. 
We used the following formula:
\begin{equation}
{\rm A_{TST}}=\frac{k_b T}{h}\frac{q^{\ddagger}}{q_{ads}}= \frac{k_b T}{h}  \frac{A}{\Lambda^2}  \frac{\sqrt[]{\pi}}{\sigma \,h^3} (8\, \pi^2 \,k_b\, T_{peak})^{3/2} ~\sqrt{I_x\, I_y \,I_z}
    \label{eq:TST}
\end{equation}
where $q_{ads}$ and $q^{\ddagger}$ are single-particle partition functions for the adsorbed (initial) state and the transition state, respectively, calculated at the temperature $T_{peak}$. \textit{A} is the surface area per adsorbed molecule. Here, like in other experiments or computational studies, it is fixed to 10$^{−19}$ m$^2$ (the inverse of the number of surface sites per unit area). $I_x$, $I_y$, and $I_z$ are the principal moments of inertia for the rotation of the particle, obtained by diagonalizing the inertia tensor of the free acetaldehyde. 
They are fixed to $I_x$ = 52.46, $I_y$ = 46.86, $I_z$ = 8.78 amu \AA$^2$. 
The symmetry factor, $\sigma$, is the number of different but indistinguishable rotational configurations of the particle and it is fixed to 1 in the case of acetaldehyde (as it belongs to the C$_s$ symmetry point group). 
The thermal wavelength of the molecule, $\Lambda$, depends on its atomic mass (44 amu) and on the peak of the desorption energy (${T_{peak}}$=115 K), and it is 24.5 pm. 
By using these parameters, ${\rm A_{TST}}$=$1.07\times 10^{18}$  s$^{-1}$. 
We emphasize, however, that ${\rm A_{TST}}$ is a function of ${T_{peak}}$. 
As shown in Fig.\ref{fig:Famille TPD CH3CHO SUR HOPG}, the temperature of the peak depends on the surface coverage. 
For example, when using ${T_{peak}}$=108 K, ${\rm A_{TST}}$ is equal to $8.58\times 10^{17}$ s$^{-1}$, while fixing ${ T_{peak}}$=117 K then ${\rm A_{TST}}$=$1.13\times10^{18}$ s$^{-1}$. 
In this study, we choose ${\rm A_{TST}}$=$1.1 \times10^{18}$ s$^{-1}$, which is the average value. 

\vspace{0.3cm}
\noindent
\textit{The astrochemical modelling point of view:} 
Often, the pre-exponential factor is calculated using the harmonic oscillator approximation introduced by \cite{Hasegawa1992} and here called ${\rm A_{HH}}$:
\begin{equation}
    {\rm A_{HH}} = \left( \frac{2~n_s~\rm BE}{\pi^2~m} \right)^{1/2}
    \label{eq:HH-pre-exponential}
\end{equation}
where $n_s$ is the surface density of sites and $m$ is the mass of the adsorbed species.
Using this approximation, which does not take into account the fact that the molecule adsorbed does no\textbf{t} rotate whereas the desorbed one does, we find ${\rm A_{HH}}=1.3\times 10^{12}$ s$^{-1}$ if we choose an arbitrary value of BE=4500 K. 
In addition, in the \cite{Hasegawa1992} formulae, the value of ${\rm A_{HH}}$ depends exclusively on the BE of the adsorbate and is calculated directly from it, so that, in the models, there is a unique parameter, BE, and not the  A and BE pair, as determined in experiments.
We stress that the chosen pre-exponential factor A can strongly affect the BE value determined using the TPD technique, since they are actually coupled (see Eq. \ref{eq:eq1}) and are somewhat degenerated (monotonic). 
To compare  the values obtained with different pre-exponential factors A, we can re-scale the BE values using ${\rm A_{HH}}$ and the following formula:
\begin{equation}
{\rm BE_{HH}}={\rm BE_{TST}}-T_{peak} ~ln({\rm A_{TST}/A_{HH}})
\label{eq3}
\end{equation}
where ${\rm BE_{HH}}$ and ${\rm A_{HH}}$ (${\rm BE_{TST}}$ and ${\rm A_{TST}}$) are the pair BE and pre-exponential factor calculated with the harmonic oscillator approximation of \cite{Hasegawa1992} or the TST proposed by \cite{tait2005b}.

\vspace{0.3cm}
\noindent
\textit{The theoretical point of view:}
A full quantum mechanical (QM) evaluation of the pre-exponential factor can be derived by the statistical mechanical treatment as in the transition state theory, considering  the availability of the harmonic frequencies for all involved systems. 
The QM pre-exponential factor $\nu_{TST}(T)$ is related to the $A_{TST}$ of Eq. \ref{eq:TST} by:
\begin{equation}\label{eq:apprx_pre_exponential}
    \nu_{TST}(T) = A_{TST} \underbrace{\frac{{^\ddagger q_{vib}(M)} {^\ddagger q_{vib}(S)}}{q_{vib}(C)}}_{\textcolor{red}{\mathbf{q_{vib}^{TST}}}}\; .
\end{equation}

Therefore the two equations coincides when the $q_{vib}^{TST}$ ratio is equal to unity. 
The application of the full TST treatment to acetaldehyde, limited to one representative case related to the P-ice model (with a BH(0) of 60.3 kJ/mol, image 6 panel A of the paper), gave the ratio between the full TST treatment and the \citet{tait2005b} prefactor close to unity for temperatures $\leq$ 50 K, decreasing by ~1 and 2 orders of magnitude at 100 K and 200 K, respectively. Therefore, considering the desorption temperature of 107 K for acetaldehyde, the \citet{tait2005b} approximation is relatively good within around one order of magnitude of the value of $\approx$ 10$^{18}$ s$^{-1}$ while being much simpler to apply.

Another advantage of adopting Eq. \ref{eq:TST} over the full QM TST one is its explicit dependence on $T_{peak}$.
For these reasons, in the following, we will adopt Eq. \ref{eq:TST} for the calculation of the pre-exponential factor.

\vspace{0.3cm}
\noindent
\textit{Comparison with previous experimental estimates:}
Now we have a tool to compare BE provided by different values of pre-exponential factors. 
\cite{Corazzi2021} recently reported an experimental value of BE=3100 K of acetaldehyde co-deposited with water on a substrate made of micrometer-sized olivine grains. 
It obviously differs from the measured values here and it is lesser than any calculated value. 
However, these authors used a fixed value of A=10$^{12}$ s$^{-1}$. 
To be able to compare with our results, we use Eq. \ref{eq3} and find that the couple BE=3079 K and A=10$^{12}s^{-1}$ corresponds to the couple BE=4680 K and A=$1.1 \times 10^{18}$ s$^{-1}$, that is exactly what has been found here for the multilayer energy of acetaldehyde. 
This example shows us how important is to take into account not only the BEs, but also the pre-exponential factor that allows us to calculate a desorption flux, to reproduce experiments or to simulate the desorption in ISM conditions.

\subsection{Comparison of experimental and theoretical results}\label{subsec:comparison}

The aim of this section is to compare the experimental and theoretical approaches. 
We will do that for the pure acetaldehyde, crystalline and amorphous water ices, separately.

\subsubsection{Pure acetaldehyde ice}
By comparing theory with experiments, in general, we compare a static and "uni-molecular" calculation with a measurement that is both dynamic and averaged over a large population and a large number of situations.
We will first compare the case of the pure acetaldehyde ice.
Here, experiments provide a BE of 4650 K, while calculations dealing with the ideal model of a molecule above the (010) acetaldehyde surface gives a BE of 2650 K. 
Given the robustness of the measurement \citep[confirmed by][]{Corazzi2021}, one would conclude that what is measured does not correspond to what is calculated. 
In reality, when a molecular film is heated, it will constantly reorganise itself, and the situation of a molecule on top of the surface is not the most energetically stable configuration.
The other extreme case corresponds to the calculation of the BE(0) of one acetaldehyde molecule extracted from the surface slab, which is 5692 K (see above). 
We think that this last computed value is probably overestimated with respect to the experimental BE, as we only focused on the process occurring at the rather stable (010) surface without considering more defective surfaces, from which the desorption can occur from kink and edges leaving the molecule less engaged with the underneath layers. 

As a first conclusion, therefore, it can be said that the present calculations provide good high and low boundary values to the experimental results, but that the dynamic (in the case of experiments) versus the static (in the case of calculations) nature can make any direct comparison somewhat misleading.

\subsubsection{CH$_3$CHO desorbing from crystalline water ice}
Although we provided two BE values because of the use of two different unit cell sizes (i.e., 1x1 and 2x1 supercell), here, for the sake of comparison with experiments, we take the value of BE=5194 K obtained using the 2x1 supercell, as it is more realistic.

In experiments, we find a distribution of BEs. 
It is not a bias of the method, as it was possible to derive a unique value of the BE for the HOPG surface. 
Actually, in experiments, even for a PCI, the surface is disordered. 
Indeed, the water molecules at the outermost positions of the surface almost randomly alternate dH pointing outside or inside the solid. 
Therefore, even for crystalline water ice, distributions of BE are usually measured whatever is the adsorbate (\cite{amiaud2007,Noble2012,Nguyen2018}. 
The calculations, made on a perfectly regular or periodic crystal, are not able to reproduce this disorder. 

In experiments, the largest population is found for BE=4650 K and it corresponds to the BE of an acetaldehyde multilayer. 
This  means that CH$_3$CHO molecules prefer to rearrange themselves, probably creating some sort of small clusters, rather than spreading over the surface. 
In other words, acetaldehyde do not wet properly the water ice surface. 
The low coverage values, which correspond more to the calculations made for a unique molecule, lay between 4950 and 5700 K. 
This is in excellent agreement with the calculated value of BE=5194 K. 
We note that population at 5700 K could also correspond to defaults or steps in the poly-crystalline assembly.

\subsubsection{CH$_3$CHO desorbing from amorphous water ice}
Both experiments and calculations show a broad distribution of BEs. Calculations demonstrates that, compared to the crystalline case, some sites are more energetically favorable (two over nine), one is about equal, and the other six are less favorable sites for adsorption. 
There is no doubt that such less bounded sites exist, but in experiments they will not be populated, as molecules reorganise during the heating phase and tend to occupy the available most favourable adsorption sites. 
This has been well experimented and documented as the "filling behaviour", for relatively volatile substances \citep{Kimmel2001,Dulieu2005}. 
In the experiments, for all adsorbates, the minimum BE measurable is set by the limit of the BE of the multilayer. This is why we observe that half of the population of acetaldehyde desorbs from sites with a BE very close to that of pure CH$_3$CHO films. 
Indeed, this accumulation at the lower part of the BE distribution corresponds, in the calculations, to the six (out of nine) sites that have a BE lower than 4650 K (the multilayer limit). 
Once again, we find here that the theoretical calculations are in very good agreement with what is obtained experimentally. 
They show perfectly the extent of the distribution (up to 6000 K), and propose a good sampling of sites. 
Finally, we point out here that we do not observe any co-desorption of acetaldehyde with water. 
All the acetaldehyde desorbs prior to the water. This is consistent with the rather low values of BE calculated for the interaction with water, lower than the multilayer BE measurements of the acetaldehyde film alone. This molecule would somehow have a low hydrophilic character.

\subsection{Comparison of the acetaldehyde BE with those of other important interstellar molecules}\label{subsec:BE-comparison}

The detailed calculations and measurements of the acetaldehyde BEs on water ices show a consistent but complex picture. 
First, as already reported in other works \citep[e.g.][]{Dulieu2005,Ferrero2020}, BE on an icy surface is not a single value, but there is rather a distribution of BEs caused by the different sites to which acetaldehyde can be adsorbed. 
Second, using the BE values without paying attention to its associated pre-exponential factor can lead to inconsistencies, if not mistakes. 
That being said, most of the astrochemical models use a single value of BE and the ${\rm A_{HH}}$ pre-exponential factor, recalculated from it following Eq. \ref{eq:HH-pre-exponential}, which is not a guarantee of correctness.

Another question is: which single value of BE to choose from the observed distribution? 
To choose a single value in the distribution, one needs to know the nature of the ice, for instance, if the surface will be fully or partly covered. 
Acetaldehyde is a frequently observed molecule in the gas phase, but it has not been detected yet in the ice mantles because its concentration is probably low (but see also the discussion in the Introduction). 
Therefore, the coverage of this molecule on the surface of icy grains should remain low and the low coverage side of the distribution shall be preferred. 
Moreover, the interval of values that we propose overlaps the BE extrapolated by \cite{Wakelam2017BE} (5400 K), which turns out to be correct if one adopts the pre-exponential factor of A=$1.1 \times 10^{18}$ s$^{-1}$.
However, if the astrochemical model uses the harmonic oscillator approximation and a value of ${\rm A_{HH}} \sim 10^{12}$ s$^{-1}$, a value around 3800 K shall be preferred.

The advantage to use the couple BE=5400 K and A=$1.1 \times 10^{18}$ s$^{-1}$ is that one can then compare this BE with that derived by the theoretical calculations and to the already published BE values of other molecules \citep{Minissale2022}. 
As an example, CH$_3$CHO can be compared to H$_2$CO, which seems to have a similar desorption behaviour, exhibiting a non co-desorption with water \citep[][]{Noble2012b}. 
The values for H$_2$CO are: BE=4117 K (A=$8.29 \times 10^{16}$ s$^{-1}$). 
On the contrary, both ethanol and methanol exhibit co-desorption with water and have larger BEs, BE=7000 K (A=$3.89 \times 10^{18}$ s$^{-1}$) \citep{Dulieu2022} and BE=6621 K (A=$3.18 \times 10^{17}$ s$^{-1}$) \citep{Bahr2008,Minissale2022}, respectively. 
Finally, formamide (NH$_2$CHO) has a refractory behaviour with respect to water. 
It desorbs at higher temperature from bare surfaces because of its high BE=9561 K (A=$3.69 \times 10^{18}$ s$^{-1}$) \citep{Chaabouni2018,Minissale2022}. 
One obtains the same substantial result when considering the theoretical BEs calculated by \cite{Ferrero2020}, although they are BE distributions.

\subsection{Astrochemical implications}
From an astrochemical point of view, two points are particularly relevant.
The first point regards the presence of acetaldehyde in cold ($\sim10$ K) objects \citep[e.g.][]{Bacmann2012,vastel2014,Scibelli2020,Zhou2022}.
The relatively large BE (4800--6000 K) makes it difficult to explain the presence of gaseous acetaldehyde if formed on the grain-surfaces, and would rather favor a gas-phase formation.
Of course, this may just move the problem of the presence of the reactants needed to synthesize acetaldehyde from them, namely ethyl radical and/or ethanol. 
Non-thermal mechanisms could be at play such as cosmic ray bombardment that would not be chemically selective (\cite{Dartois2019}), whereas the chemical desorption initiated by hydrogenation on the grain surface may have different efficiencies depending on the molecule \citep{Minissale2016}.

As written in the Introduction, a the BE of a given species determines at what temperature the species remains adsorbed or goes to the gas-phase.
In hot cores/corinos, several species, including acetaldehyde, have a jump in the abundance in the region where the dust temperature reaches about 100 K.
This is believed to be caused by the sublimation of the frozen water \citep[e.g.][]{Charnley1992,Ceccarelli2000a,Jaber2014}, which is the major constituent of the icy grain mantles \citep[e.g.][]{Boogert2015}.
However, the analysis of the emission lines of different species and their spatial distributions sometimes suggest that there is a differentiation in the sublimation of different species \citep[e.g.][]{Manigand2020,Bianchi2022}.

The recent study by \cite{Bianchi2022} definitively shows a chemical differentiation between the two hot corinos of the SVS13 protobinary system.
Specifically, the analysis of the high spatial resolution ALMA observations leads to the conclusion that gaseous acetaldehyde and formamide (NH$_2$CHO) are distributed in an onion-like structure of the hot corino, with formamide becoming abundant in a warmer region with respect to the acetaldehyde region.
A natural explanation of this behavior is that species with larger BE emit lines (mostly) in more compact regions, i.e., the region corresponding to their sublimation temperature, rather than the water sublimation front, because the large densities ($\geq 10^8$ cm$^{-3}$) make those species freeze-out back very quickly. 
Therefore, given their respective BEs (see \S ~\ref{subsec:BE-comparison}), acetaldehyde is emitted in a region more extended and colder than that of formamide. 

More in general, by considering their respective BEs, formaldehyde and acetaldehyde would desorb before water, ethanol and methanol with water, but formamide has to wait for higher average temperatures, given its much higher BE.
Our new estimates of the acetaldehyde BE, slightly lower than water, would suggest that acetaldehyde should be found approximately in the regions where water and methanol are present too, which is what has been so far found \citep{Bianchi2022}.
On the contrary, formamide has definitively a larger BE range, so that we predict formamide to originate in a hotter region than that of acetaldehyde, which is indeed what is observed in the few cases where a similar analysis has been carried out \citep[e.g.][]{Csengeri2019,Okoda2021,Bianchi2022}.

In summary, with the high sensitivity of the new present facilities, such as ALMA and NOEMA, very likely studies revealing a differentiation in the chemical species in hot cores/corinos will become more an more available.
We need to be prepared with studies similar to the one reported here, where the BEs of more complex organic molecules are estimated, so that we can appropriately interpret the astronomical observations.
In turn, those observations can help to validate our methods, for which so much uncertainty still persist.

\section{Conclusions}\label{sec:conclusions}

We presented TPD experiments and quantum chemical computations of the acetaldehyde BE on a pure acetaldehyde ice, and on crystalline and amorphous water ices.
The main conclusions of this work are the following:
\begin{enumerate}
    \item The experiments indicate that the acetaldehyde BE has a distribution of values.
    In the acetaldehyde ice, the BE has two main energy values, at about 4650 and 5250 K, with the peak around 5250 K being about ten times more frequent.
    In crystalline water ice, the BE ranges between about 4600 and 5700 K, with a peak around 4600--4800 K.
    In amorphous water ice, the BE ranges between about 4600 and 5900 K, with a peak around 4800 K.
    The weighted average in the three cases are 5150, 5080 and 4990 K, respectively.
    \item The theoretical calculations give BE values on the different sites of the used ice model.
    BEs spread over 5194 and 7253 K in the case of crystalline ice, and 2809 to 6038 K in the case of amorphous ice.
    In the pure acetaldehyde ice, two extreme values of 2650 and 5692 K are obtained.
    \item We discussed and showed the importance of the pre-exponential factor when deriving, comparing and using BE derived from experiments and theoretical calculations. Remarkably, when the correct pre-exponential factor is used with the derived BE, experiments and theory are in fair good agreement.
    \item A comparison of the the derived acetaldehyde BEs with those of other important species shows that acetaldehyde would desorb at temperatures lower than those at which water desorbs and even lower temperatures with respect to formamide.
    \item The large acetaldehyde BE challenges the explanation for its gas-phase presence in cold ($\sim 10$ K) astronomical objects, especially if it is formed on the grain surfaces. 
    The problem may be alleviated if it is formed by gas-phase reactions.
    \item In hot cores/corinos, the measured and computed BE is in agreement with the observations of acetaldehyde originating in regions colder than formamide, whose BE is larger.
\end{enumerate}
Finally, our study shows the importance to extend the methodology adopted here to other molecules of astrochemical interest, such as ethanol or formamide, which are nowadays routinely observed in astronomical objects and that show a spatial segregation probably due to their different BE.

\section*{Acknowledgements}

This project has received funding within the European Union’s Horizon 2020 research and innovation programme from the European Research Council (ERC) for the projects ``The Dawn of Organic Chemistry” (DOC), grant agreement No 741002 and ``Quantum Chemistry on Interstellar Grains” (QUANTUMGRAIN), grant agreement No 865657, and from the Marie Sklodowska-Curie for the project ``Astro-Chemical Origins” (ACO), grant agreement No 811312.  
This work was supported by the Agence Nationale de la recherche (ANR) SIRC project (Grant ANR-SPV202448 2020-2024).
CC wishes to thank Eleonora Bianchi and Lorenzo Tinacci for fruitful discussions on the chemical segregation observed in hot corinos and the evaluation of the theoretical binding energy, respectively.
This work has also been funded by the European Research Council (ERC) for the Starting Grant "DustOrigin”, hold by Prof. Ilse De Looze, University of Ghent, grant agreement ID: 851622. 
AR is indebted to the \textit{Ramón y Cajal} programme.


\section*{Data availability}

The data underlying this article are available in Zenodo at \url{https://zenodo.org/communities/aco-astro-chemical-origins/?page=1&size=20}



\bibliographystyle{mnras}
\bibliography{CC_bibtex} 

\begin{thebibliography}{}
\makeatletter
\relax
\def\mn@urlcharsother{\let\do\@makeother \do\$\do\&\do\#\do\^\do\_\do\%\do\~}
\def\mn@doi{\begingroup\mn@urlcharsother \@ifnextchar [ {\mn@doi@}
  {\mn@doi@[]}}
\def\mn@doi@[#1]#2{\def\@tempa{#1}\ifx\@tempa\@empty \href
  {http://dx.doi.org/#2} {doi:#2}\else \href {http://dx.doi.org/#2} {#1}\fi
  \endgroup}
\def\mn@eprint#1#2{\mn@eprint@#1:#2::\@nil}
\def\mn@eprint@arXiv#1{\href {http://arxiv.org/abs/#1} {{\tt arXiv:#1}}}
\def\mn@eprint@dblp#1{\href {http://dblp.uni-trier.de/rec/bibtex/#1.xml}
  {dblp:#1}}
\def\mn@eprint@#1:#2:#3:#4\@nil{\def\@tempa {#1}\def\@tempb {#2}\def\@tempc
  {#3}\ifx \@tempc \@empty \let \@tempc \@tempb \let \@tempb \@tempa \fi \ifx
  \@tempb \@empty \def\@tempb {arXiv}\fi \@ifundefined
  {mn@eprint@\@tempb}{\@tempb:\@tempc}{\expandafter \expandafter \csname
  mn@eprint@\@tempb\endcsname \expandafter{\@tempc}}}

\bibitem[\protect\citeauthoryear{{Amiaud}, {Dulieu}, {Fillion}, {Momeni}  \&
  {Lamaire}}{{Amiaud} et~al.}{2007}]{amiaud2007}
{Amiaud} L.,  {Dulieu} F.,  {Fillion} J.~H.,  {Momeni} A.,   {Lamaire} J.~L.,
  2007, \mn@doi [J.Chem.Phys.] {10.1063/1.2746323}, \href
  {https://ui.adsabs.harvard.edu/abs/2007JChPh.127n4709A/abstract} {127, 12}

\bibitem[\protect\citeauthoryear{{Bacmann}, {Taquet}, {Faure}, {Kahane}  \&
  {Ceccarelli}}{{Bacmann} et~al.}{2012}]{Bacmann2012}
{Bacmann} A.,  {Taquet} V.,  {Faure} A.,  {Kahane} C.,   {Ceccarelli} C.,
  2012, \mn@doi [\aap] {10.1051/0004-6361/201219207}, \href
  {https://ui.adsabs.harvard.edu/abs/2012A&A...541L..12B} {541, L12}

\bibitem[\protect\citeauthoryear{Bahr, Toubin  \& Kempter}{Bahr
  et~al.}{2008}]{Bahr2008}
Bahr S.,  Toubin C.,   Kempter V.,  2008, \mn@doi [J. Chem. Phys.]
  {10.1063/1.2901970}, 128, 134712

\bibitem[\protect\citeauthoryear{Becke}{Becke}{1993}]{Becke1993}
Becke A.~D.,  1993, J. Chem. Phys., 98, 1372

\bibitem[\protect\citeauthoryear{{Bennett}, {Jamieson}, {Osamura}  \&
  {Kaiser}}{{Bennett} et~al.}{2005}]{Bennett2005}
{Bennett} C.~J.,  {Jamieson} C.~S.,  {Osamura} Y.,   {Kaiser} R.~I.,  2005,
  \mn@doi [\apj] {10.1086/429119}, \href
  {https://ui.adsabs.harvard.edu/abs/2005ApJ...624.1097B} {624, 1097}

\bibitem[\protect\citeauthoryear{{Bianchi} et~al.,}{{Bianchi}
  et~al.}{2019}]{Bianchi2019}
{Bianchi} E.,  et~al., 2019, \mn@doi [\mnras] {10.1093/mnras/sty2915}, \href
  {https://ui.adsabs.harvard.edu/abs/2019MNRAS.483.1850B} {483, 1850}

\bibitem[\protect\citeauthoryear{{Bianchi}, {L{\'o}pez-Sepulcre}, {Ceccarelli},
  {Codella}, {Podio}, {Bouvier}  \& {Enrique-Romero}}{{Bianchi}
  et~al.}{2022}]{Bianchi2022}
{Bianchi} E.,  {L{\'o}pez-Sepulcre} A.,  {Ceccarelli} C.,  {Codella} C.,
  {Podio} L.,  {Bouvier} M.,   {Enrique-Romero} J.,  2022, arXiv e-prints,
  \href {https://ui.adsabs.harvard.edu/abs/2022arXiv220303412B} {p.
  arXiv:2203.03412}

\bibitem[\protect\citeauthoryear{{Biver} et~al.,}{{Biver}
  et~al.}{2021}]{Biver2021}
{Biver} N.,  et~al., 2021, \mn@doi [\aap] {10.1051/0004-6361/202040125}, \href
  {https://ui.adsabs.harvard.edu/abs/2021A&A...648A..49B} {648, A49}

\bibitem[\protect\citeauthoryear{{Blake}, {Sutton}, {Masson}  \&
  {Phillips}}{{Blake} et~al.}{1987}]{Blake1987}
{Blake} G.~A.,  {Sutton} E.~C.,  {Masson} C.~R.,   {Phillips} T.~G.,  1987,
  \mn@doi [\apj] {10.1086/165165}, \href
  {https://ui.adsabs.harvard.edu/abs/1987ApJ...315..621B} {315, 621}

\bibitem[\protect\citeauthoryear{{Boogert}, {Gerakines}  \&
  {Whittet}}{{Boogert} et~al.}{2015}]{Boogert2015}
{Boogert} A.~C.~A.,  {Gerakines} P.~A.,   {Whittet} D. C.~B.,  2015, \mn@doi
  [\araa] {10.1146/annurev-astro-082214-122348}, \href
  {https://ui.adsabs.harvard.edu/abs/2015ARA&A..53..541B} {53, 541}

\bibitem[\protect\citeauthoryear{Bushlanov, Blatov  \& Oganov}{Bushlanov
  et~al.}{2019}]{bushlanov2019topology}
Bushlanov P.~V.,  Blatov V.~A.,   Oganov A.~R.,  2019, Comput. Phys. Commun.,
  236, 1

\bibitem[\protect\citeauthoryear{Caldeweyher, Bannwarth  \& Grimme}{Caldeweyher
  et~al.}{2017}]{caldeweyher2017extension}
Caldeweyher E.,  Bannwarth C.,   Grimme S.,  2017, J. Chem. Phys., 147, 034112

\bibitem[\protect\citeauthoryear{Caldeweyher, Ehlert, Hansen, Neugebauer,
  Spicher, Bannwarth  \& Grimme}{Caldeweyher
  et~al.}{2019}]{caldeweyher2019generally}
Caldeweyher E.,  Ehlert S.,  Hansen A.,  Neugebauer H.,  Spicher S.,  Bannwarth
  C.,   Grimme S.,  2019, J. Chem. Phys., 150, 154122

\bibitem[\protect\citeauthoryear{{Cazaux}, {Tielens}, {Ceccarelli}, {Castets},
  {Wakelam}, {Caux}, {Parise}  \& {Teyssier}}{{Cazaux}
  et~al.}{2003}]{Cazaux2003}
{Cazaux} S.,  {Tielens} A.~G.~G.~M.,  {Ceccarelli} C.,  {Castets} A.,
  {Wakelam} V.,  {Caux} E.,  {Parise} B.,   {Teyssier} D.,  2003, \mn@doi
  [\apjl] {10.1086/378038}, \href
  {https://ui.adsabs.harvard.edu/abs/2003ApJ...593L..51C} {593, L51}

\bibitem[\protect\citeauthoryear{{Ceccarelli}, {Loinard}, {Castets}, {Tielens}
  \& {Caux}}{{Ceccarelli} et~al.}{2000}]{Ceccarelli2000a}
{Ceccarelli} C.,  {Loinard} L.,  {Castets} A.,  {Tielens} A.~G.~G.~M.,   {Caux}
  E.,  2000, \aap, \href
  {https://ui.adsabs.harvard.edu/abs/2000A&A...357L...9C} {357, L9}

\bibitem[\protect\citeauthoryear{Chaabouni, Diana, Nguyen  \& Dulieu}{Chaabouni
  et~al.}{2018}]{Chaabouni2018}
Chaabouni H.,  Diana S.,  Nguyen T.,   Dulieu F.,  2018, \mn@doi [\aap]
  {10.1051/0004-6361/201731006}, 612, A47

\bibitem[\protect\citeauthoryear{{Chaabouni}, {Minissale}, {Baouche}  \&
  {Dulieu}}{{Chaabouni} et~al.}{2020}]{chaabouni2020}
{Chaabouni} H.,  {Minissale} M.,  {Baouche} S.,   {Dulieu} F.,  2020, \HAL

\bibitem[\protect\citeauthoryear{{Chahine} et~al.,}{{Chahine}
  et~al.}{2022}]{Chahine2022}
{Chahine} L.,  et~al., 2022, \mn@doi [\aap] {10.1051/0004-6361/202141811},
  \href {https://ui.adsabs.harvard.edu/abs/2022A&A...657A..78C} {657, A78}

\bibitem[\protect\citeauthoryear{{Charnley}}{{Charnley}}{2004}]{Charnley2004}
{Charnley} S.~B.,  2004, \mn@doi [Adv. Space Res.] {10.1016/j.asr.2003.08.005},
  \href {https://ui.adsabs.harvard.edu/abs/2004AdSpR..33...23C} {33, 23}

\bibitem[\protect\citeauthoryear{{Charnley}, {Tielens}  \& {Millar}}{{Charnley}
  et~al.}{1992}]{Charnley1992}
{Charnley} S.~B.,  {Tielens} A.~G.~G.~M.,   {Millar} T.~J.,  1992, \mn@doi
  [\apjl] {10.1086/186609}, \href
  {https://ui.adsabs.harvard.edu/abs/1992ApJ...399L..71C} {399, L71}

\bibitem[\protect\citeauthoryear{{Chuang}, {Fedoseev}, {Scir{\`e}}, {Baratta},
  {J{\"a}ger}, {Henning}, {Linnartz}  \& {Palumbo}}{{Chuang}
  et~al.}{2021}]{Chuang2021}
{Chuang} K.~J.,  {Fedoseev} G.,  {Scir{\`e}} C.,  {Baratta} G.~A.,  {J{\"a}ger}
  C.,  {Henning} T.,  {Linnartz} H.,   {Palumbo} M.~E.,  2021, \mn@doi [\aap]
  {10.1051/0004-6361/202140780}, \href
  {https://ui.adsabs.harvard.edu/abs/2021A&A...650A..85C} {650, A85}

\bibitem[\protect\citeauthoryear{{Codella} et~al.,}{{Codella}
  et~al.}{2018}]{Codella2018}
{Codella} C.,  et~al., 2018, \mn@doi [\aap] {10.1051/0004-6361/201832765},
  \href {https://ui.adsabs.harvard.edu/abs/2018A&A...617A..10C} {617, A10}

\bibitem[\protect\citeauthoryear{{Codella} et~al.,}{{Codella}
  et~al.}{2020}]{codella2020}
{Codella} C.,  et~al., 2020, \mn@doi [\aap] {10.1051/0004-6361/201936725},
  \href {https://ui.adsabs.harvard.edu/abs/2020A&A...635A..17C} {635, A17}

\bibitem[\protect\citeauthoryear{Congiu, Chaabouni, Laffon, Parent, Baouche  \&
  Dulieu}{Congiu et~al.}{2012}]{Congiu2012}
Congiu E.,  Chaabouni H.,  Laffon C.,  Parent P.,  Baouche S.,   Dulieu F.,
  2012, \mn@doi [J. Chem. Phys.] {10.1063/1.4738895}, 137, 054713

\bibitem[\protect\citeauthoryear{{Corazzi}, {Brucato}, {Poggiali}, {Podio},
  {Fedele}  \& {Codella}}{{Corazzi} et~al.}{2021}]{Corazzi2021}
{Corazzi} M.~A.,  {Brucato} J.~R.,  {Poggiali} G.,  {Podio} L.,  {Fedele} D.,
  {Codella} C.,  2021, \mn@doi [\apj] {10.3847/1538-4357/abf6d3}, \href
  {https://ui.adsabs.harvard.edu/abs/2021ApJ...913..128C} {913, 128}

\bibitem[\protect\citeauthoryear{{Crovisier}, {Bockel{\'e}e-Morvan}, {Colom},
  {Biver}, {Despois}, {Lis}  \& {Teamtarget-of-opportunity radio observations
  of comets}}{{Crovisier} et~al.}{2004}]{Crovisier2004}
{Crovisier} J.,  {Bockel{\'e}e-Morvan} D.,  {Colom} P.,  {Biver} N.,  {Despois}
  D.,  {Lis} D.~C.,   {Teamtarget-of-opportunity radio observations of comets}
  2004, \mn@doi [\aap] {10.1051/0004-6361:20035688}, \href
  {https://ui.adsabs.harvard.edu/abs/2004A&A...418.1141C} {418, 1141}

\bibitem[\protect\citeauthoryear{{Csengeri}, {Belloche}, {Bontemps},
  {Wyrowski}, {Menten}  \& {Bouscasse}}{{Csengeri} et~al.}{2019}]{Csengeri2019}
{Csengeri} T.,  {Belloche} A.,  {Bontemps} S.,  {Wyrowski} F.,  {Menten} K.~M.,
    {Bouscasse} L.,  2019, \mn@doi [\aap] {10.1051/0004-6361/201935226}, \href
  {https://ui.adsabs.harvard.edu/abs/2019A&A...632A..57C} {632, A57}

\bibitem[\protect\citeauthoryear{Dartois, Chabot, {Id Barkach}, Rothard,
  Aug{\'{e}}, Agnihotri, Domaracka  \& Boduch}{Dartois
  et~al.}{2019}]{Dartois2019}
Dartois E.,  Chabot M.,  {Id Barkach} T.,  Rothard H.,  Aug{\'{e}} B.,
  Agnihotri A.~N.,  Domaracka A.,   Boduch P.,  2019, \mn@doi [\aap]
  {10.1051/0004-6361/201834787}, 627, A55

\bibitem[\protect\citeauthoryear{{De Simone} et~al.,}{{De Simone}
  et~al.}{2020}]{DeSimone2020_acet}
{De Simone} M.,  et~al., 2020, \mn@doi [\aap] {10.1051/0004-6361/201937004},
  \href {https://ui.adsabs.harvard.edu/abs/2020A&A...640A..75D} {640, A75}

\bibitem[\protect\citeauthoryear{Doná, Brandenburg  \& Civalleri}{Doná
  et~al.}{2019}]{PBE0sol-3c}
Doná L.,  Brandenburg J.~G.,   Civalleri B.,  2019, \mn@doi [J. Chem. Phys.]
  {10.1063/1.5123627}, 151, 121101

\bibitem[\protect\citeauthoryear{Dovesi et~al.,}{Dovesi
  et~al.}{2018}]{dovesi2018quantum}
Dovesi R.,  et~al., 2018, WIREs Comput. Mol. Sci., 8, e1360

\bibitem[\protect\citeauthoryear{{Drozdovskaya}, {van Dishoeck}, {Rubin},
  {J{\o}rgensen}  \& {Altwegg}}{{Drozdovskaya} et~al.}{2019}]{Drozdovskaya2019}
{Drozdovskaya} M.~N.,  {van Dishoeck} E.~F.,  {Rubin} M.,  {J{\o}rgensen}
  J.~K.,   {Altwegg} K.,  2019, \mn@doi [\mnras] {10.1093/mnras/stz2430}, \href
  {https://ui.adsabs.harvard.edu/abs/2019MNRAS.490...50D} {490, 50}

\bibitem[\protect\citeauthoryear{Dulieu, Amiaud, Baouche, Momeni, Fillion  \&
  Lemaire}{Dulieu et~al.}{2005}]{Dulieu2005}
Dulieu F.,  Amiaud L.,  Baouche S.,  Momeni A.,  Fillion J.-H.,   Lemaire J.,
  2005, \mn@doi [Chem. Phys. Lett.] {10.1016/j.cplett.2005.01.044}, 404, 187

\bibitem[\protect\citeauthoryear{Dulieu, Vitorino  \& Minissale}{Dulieu
  et~al.}{2022}]{Dulieu2022}
Dulieu F.,  Vitorino J.,   Minissale M.,  2022, {Private communication}

\bibitem[\protect\citeauthoryear{Elstner, Porezag, Jungnickel, Elsner, Haugk,
  Frauenheim, Suhai  \& Seifert}{Elstner et~al.}{1998}]{elstner1998self}
Elstner M.,  Porezag D.,  Jungnickel G.,  Elsner J.,  Haugk M.,  Frauenheim T.,
   Suhai S.,   Seifert G.,  1998, Phys. Rev. B, 58, 7260

\bibitem[\protect\citeauthoryear{{Enrique-Romero}, {Rimola}, {Ceccarelli}  \&
  {Balucani}}{{Enrique-Romero} et~al.}{2016}]{Enrique-Romero2016}
{Enrique-Romero} J.,  {Rimola} A.,  {Ceccarelli} C.,   {Balucani} N.,  2016,
  \mn@doi [\mnras] {10.1093/mnrasl/slw031}, \href
  {https://ui.adsabs.harvard.edu/abs/2016MNRAS.459L...6E} {459, L6}

\bibitem[\protect\citeauthoryear{{Enrique-Romero}, {Rimola}, {Ceccarelli},
  {Ugliengo}, {Balucani}  \& {Skouteris}}{{Enrique-Romero}
  et~al.}{2019}]{Enrique-Romero2019}
{Enrique-Romero} J.,  {Rimola} A.,  {Ceccarelli} C.,  {Ugliengo} P.,
  {Balucani} N.,   {Skouteris} D.,  2019, \mn@doi [ACS Earth Space Chem.]
  {10.1021/acsearthspacechem.9b00156}, \href
  {https://ui.adsabs.harvard.edu/abs/2019ECS.....3.2158E} {3, 2158}

\bibitem[\protect\citeauthoryear{{Enrique-Romero} et~al.,}{{Enrique-Romero}
  et~al.}{2020}]{Enrique-Romero2020}
{Enrique-Romero} J.,  et~al., 2020, \mn@doi [\mnras] {10.1093/mnras/staa484},
  \href {https://ui.adsabs.harvard.edu/abs/2020MNRAS.493.2523E} {493, 2523}

\bibitem[\protect\citeauthoryear{{Enrique-Romero}, {Ceccarelli}, {Rimola},
  {Skouteris}, {Balucani}  \& {Ugliengo}}{{Enrique-Romero}
  et~al.}{2021}]{Enrique-Romero2021}
{Enrique-Romero} J.,  {Ceccarelli} C.,  {Rimola} A.,  {Skouteris} D.,
  {Balucani} N.,   {Ugliengo} P.,  2021, \mn@doi [\aap]
  {10.1051/0004-6361/202141531}, \href
  {https://ui.adsabs.harvard.edu/abs/2021A&A...655A...9E} {655, A9}

\bibitem[\protect\citeauthoryear{{Fedoseev}, {Qasim}, {Chuang}, {Ioppolo},
  {Lamberts}, {van Dishoeck}  \& {Linnartz}}{{Fedoseev}
  et~al.}{2022}]{Fedoseev2022}
{Fedoseev} G.,  {Qasim} D.,  {Chuang} K.-J.,  {Ioppolo} S.,  {Lamberts} T.,
  {van Dishoeck} E.~F.,   {Linnartz} H.,  2022, \mn@doi [\apj]
  {10.3847/1538-4357/ac3834}, \href
  {https://ui.adsabs.harvard.edu/abs/2022ApJ...924..110F} {924, 110}

\bibitem[\protect\citeauthoryear{{Ferrero}, {Zamirri}, {Ceccarelli}, {Witzel},
  {Rimola}  \& {Ugliengo}}{{Ferrero} et~al.}{2020}]{Ferrero2020}
{Ferrero} S.,  {Zamirri} L.,  {Ceccarelli} C.,  {Witzel} A.,  {Rimola} A.,
  {Ugliengo} P.,  2020, \mn@doi [\apj] {10.3847/1538-4357/abb953}, \href
  {https://ui.adsabs.harvard.edu/abs/2020ApJ...904...11F} {904, 11}

\bibitem[\protect\citeauthoryear{{Fourikis}, {Sinclair}, {Robinson}, {Godfrey}
  \& {Brown}}{{Fourikis} et~al.}{1974}]{Fourikis1974}
{Fourikis} N.,  {Sinclair} M.~W.,  {Robinson} B.~J.,  {Godfrey} P.~D.,
  {Brown} R.~D.,  1974, \mn@doi [Austral. J. Phys.] {10.1071/PH740425}, \href
  {https://ui.adsabs.harvard.edu/abs/1974AuJPh..27..425F} {27, 425}

\bibitem[\protect\citeauthoryear{{Garrod} \& {Herbst}}{{Garrod} \&
  {Herbst}}{2006}]{Garrod2006}
{Garrod} R.~T.,  {Herbst} E.,  2006, \mn@doi [\aap]
  {10.1051/0004-6361:20065560}, \href
  {https://ui.adsabs.harvard.edu/abs/2006A&A...457..927G} {457, 927}

\bibitem[\protect\citeauthoryear{Gaussian09}{Gaussian09}{2009}]{gaussian0920091}
Gaussian09 R.~A.,  2009, Inc., Wallingford CT, 121, 150

\bibitem[\protect\citeauthoryear{{Gottlieb}}{{Gottlieb}}{1973}]{Gottlieb1973}
{Gottlieb} C.~A.,  1973, in {Gordon} M.~A.,  {Snyder} L.~E.,  eds, Molecules in
  the Galactic Environment. p.~181

\bibitem[\protect\citeauthoryear{Grimme, Antony, Ehrlich  \& Krieg}{Grimme
  et~al.}{2010}]{Grimme2010}
Grimme S.,  Antony J.,  Ehrlich S.,   Krieg H.,  2010, J. Chem. Phys., 132,
  154104

\bibitem[\protect\citeauthoryear{Grimme, Ehrlich  \& Goerigk}{Grimme
  et~al.}{2011}]{Grimme2011}
Grimme S.,  Ehrlich S.,   Goerigk L.,  2011, \mn@doi [J. Comput. Chem]
  {https://doi.org/10.1002/jcc.21759}, 32, 1456

\bibitem[\protect\citeauthoryear{{Guti{\'e}rrez-Quintanilla}
  et~al.,}{{Guti{\'e}rrez-Quintanilla}
  et~al.}{2021}]{Gutierrez-Quintanilla2021}
{Guti{\'e}rrez-Quintanilla} A.,  et~al., 2021, \mn@doi [\mnras]
  {10.1093/mnras/stab1850}, \href
  {https://ui.adsabs.harvard.edu/abs/2021MNRAS.506.3734G} {506, 3734}

\bibitem[\protect\citeauthoryear{{Hasegawa}, {Herbst}  \& {Leung}}{{Hasegawa}
  et~al.}{1992}]{Hasegawa1992}
{Hasegawa} T.~I.,  {Herbst} E.,   {Leung} C.~M.,  1992, \mn@doi [\apjs]
  {10.1086/191713}, \href
  {https://ui.adsabs.harvard.edu/abs/1992ApJS...82..167H} {82, 167}

\bibitem[\protect\citeauthoryear{Hourahine et~al.,}{Hourahine
  et~al.}{2020}]{hourahine2020dftb+}
Hourahine B.,  et~al., 2020, J. Chem. Phys., 152, 124101

\bibitem[\protect\citeauthoryear{{Hudson} \& {Ferrante}}{{Hudson} \&
  {Ferrante}}{2020}]{Hudson2020}
{Hudson} R.~L.,  {Ferrante} R.~F.,  2020, \mn@doi [\mnras]
  {10.1093/mnras/stz3323}, \href
  {https://ui.adsabs.harvard.edu/abs/2020MNRAS.492..283H} {492, 283}

\bibitem[\protect\citeauthoryear{{Hudson} \& {Moore}}{{Hudson} \&
  {Moore}}{1997}]{Hudson1997}
{Hudson} R.~L.,  {Moore} M.~H.,  1997, \mn@doi [\icarus]
  {10.1006/icar.1997.5678}, \href
  {https://ui.adsabs.harvard.edu/abs/1997Icar..126..233H} {126, 233}

\bibitem[\protect\citeauthoryear{{Jaber}, {Ceccarelli}, {Kahane}  \&
  {Caux}}{{Jaber} et~al.}{2014}]{Jaber2014}
{Jaber} A.~A.,  {Ceccarelli} C.,  {Kahane} C.,   {Caux} E.,  2014, \mn@doi
  [\apj] {10.1088/0004-637X/791/1/29}, \href
  {https://ui.adsabs.harvard.edu/abs/2014ApJ...791...29J} {791, 29}

\bibitem[\protect\citeauthoryear{Kimmel, Stevenson, Dohn{'a}lek, Smith  \&
  Kay}{Kimmel et~al.}{2001}]{Kimmel2001}
Kimmel G.~a.,  Stevenson K.~P.,  Dohn{'a}lek Z.,  Smith R.~S.,   Kay B.~D.,
  2001, \mn@doi [J. Chem. Phys.] {10.1063/1.1350580}, 114, 5284

\bibitem[\protect\citeauthoryear{Kresse \& Furthm{\"u}ller}{Kresse \&
  Furthm{\"u}ller}{1996a}]{kresse1996efficiency}
Kresse G.,  Furthm{\"u}ller J.,  1996a, Comput. Materials Sci., 6, 15

\bibitem[\protect\citeauthoryear{Kresse \& Furthm{\"u}ller}{Kresse \&
  Furthm{\"u}ller}{1996b}]{kresse1996efficient}
Kresse G.,  Furthm{\"u}ller J.,  1996b, Phys. Rev. B, 54, 11169

\bibitem[\protect\citeauthoryear{Kresse \& Hafner}{Kresse \&
  Hafner}{1993}]{kresse1993ab}
Kresse G.,  Hafner J.,  1993, Phys. Rev. B, 47, 558

\bibitem[\protect\citeauthoryear{Kresse \& Hafner}{Kresse \&
  Hafner}{1994}]{kresse1994ab}
Kresse G.,  Hafner J.,  1994, Phys. Rev. B, 49, 14251

\bibitem[\protect\citeauthoryear{Kresse \& Joubert}{Kresse \&
  Joubert}{1999}]{kresse1999ultrasoft}
Kresse G.,  Joubert D.,  1999, Phys. Rev. B, 59, 1758

\bibitem[\protect\citeauthoryear{{Lamberts}, {Markmeyer}, {Kolb}  \&
  {K{\"a}stner}}{{Lamberts} et~al.}{2019}]{Lamberts2019}
{Lamberts} T.,  {Markmeyer} M.~N.,  {Kolb} F.~J.,   {K{\"a}stner} J.,  2019,
  \mn@doi [ACS Earth Space Chem.] {10.1021/acsearthspacechem.9b00029}, \href
  {https://ui.adsabs.harvard.edu/abs/2019ESC.....3..958L} {3, 958}

\bibitem[\protect\citeauthoryear{{Law}, {Zhang}, {{\"O}berg},
  {Galv{\'a}n-Madrid}, {Keto}, {Liu}  \& {Ho}}{{Law} et~al.}{2021}]{Law2021}
{Law} C.~J.,  {Zhang} Q.,  {{\"O}berg} K.~I.,  {Galv{\'a}n-Madrid} R.,  {Keto}
  E.,  {Liu} H.~B.,   {Ho} P. T.~P.,  2021, \mn@doi [\apj]
  {10.3847/1538-4357/abdeb8}, \href
  {https://ui.adsabs.harvard.edu/abs/2021ApJ...909..214L} {909, 214}

\bibitem[\protect\citeauthoryear{{Le Roy} et~al.,}{{Le Roy}
  et~al.}{2015}]{LeRoy2015}
{Le Roy} L.,  et~al., 2015, \mn@doi [\aap] {10.1051/0004-6361/201526450}, \href
  {https://ui.adsabs.harvard.edu/abs/2015A&A...583A...1L} {583, A1}

\bibitem[\protect\citeauthoryear{Lee, Yang  \& Parr}{Lee et~al.}{1988}]{LYP88}
Lee C.,  Yang W.,   Parr R.~G.,  1988, \mn@doi [Phys. Rev. B]
  {10.1103/PhysRevB.37.785}, 37, 785

\bibitem[\protect\citeauthoryear{{Lee}, {Codella}, {Li}  \& {Liu}}{{Lee}
  et~al.}{2019}]{Lee2019}
{Lee} C.-F.,  {Codella} C.,  {Li} Z.-Y.,   {Liu} S.-Y.,  2019, \mn@doi [\apj]
  {10.3847/1538-4357/ab15db}, \href
  {https://ui.adsabs.harvard.edu/abs/2019ApJ...876...63L} {876, 63}

\bibitem[\protect\citeauthoryear{{Lefloch}, {Ceccarelli}, {Codella}, {Favre},
  {Podio}, {Vastel}, {Viti}  \& {Bachiller}}{{Lefloch}
  et~al.}{2017}]{Lefloch2017}
{Lefloch} B.,  {Ceccarelli} C.,  {Codella} C.,  {Favre} C.,  {Podio} L.,
  {Vastel} C.,  {Viti} S.,   {Bachiller} R.,  2017, \mn@doi [\mnras]
  {10.1093/mnrasl/slx050}, \href
  {https://ui.adsabs.harvard.edu/abs/2017MNRAS.469L..73L} {469, L73}

\bibitem[\protect\citeauthoryear{Lyakhov, Oganov, Stokes  \& Zhu}{Lyakhov
  et~al.}{2013}]{lyakhov2013new}
Lyakhov A.~O.,  Oganov A.~R.,  Stokes H.~T.,   Zhu Q.,  2013, Comput. Phys.
  Commun., 184, 1172

\bibitem[\protect\citeauthoryear{{Manigand} et~al.,}{{Manigand}
  et~al.}{2020}]{Manigand2020}
{Manigand} S.,  et~al., 2020, \mn@doi [\aap] {10.1051/0004-6361/201936299},
  \href {https://ui.adsabs.harvard.edu/abs/2020A&A...635A..48M} {635, A48}

\bibitem[\protect\citeauthoryear{{Mart{\'\i}n-Dom{\'e}nech}, {{\"O}berg}  \&
  {Rajappan}}{{Mart{\'\i}n-Dom{\'e}nech} et~al.}{2020}]{MartinDomenech2020}
{Mart{\'\i}n-Dom{\'e}nech} R.,  {{\"O}berg} K.~I.,   {Rajappan} M.,  2020,
  \mn@doi [\apj] {10.3847/1538-4357/ab84e8}, \href
  {https://ui.adsabs.harvard.edu/abs/2020ApJ...894...98M} {894, 98}

\bibitem[\protect\citeauthoryear{McElroy, Walsh, Markwick, Cordiner, Smith  \&
  Millar}{McElroy et~al.}{2013}]{mcelroy2013}
McElroy D.,  Walsh C.,  Markwick A.,  Cordiner M.,  Smith K.,   Millar T.,
  2013, \mn@doi [\aap] {10.1051/0004-6361/201220465}, 550

\bibitem[\protect\citeauthoryear{Minissale, Dulieu, Cazaux  \& Hocuk}{Minissale
  et~al.}{2016}]{Minissale2016}
Minissale M.,  Dulieu F.,  Cazaux S.,   Hocuk S.,  2016, \mn@doi [\aap]
  {10.1051/0004-6361/201525981}, 585

\bibitem[\protect\citeauthoryear{Minissale et~al.,}{Minissale
  et~al.}{2022}]{Minissale2022}
Minissale M.,  et~al., 2022, \mn@doi [ACS Earth Space Chem.]
  {10.1021/acsearthspacechem.1c00357}, 6, 597

\bibitem[\protect\citeauthoryear{Molpeceres, Kaestner, Herrero, Peláez  \&
  Maté}{Molpeceres et~al.}{2022}]{Molpeceres2022}
Molpeceres G.,  Kaestner J.,  Herrero V.,  Peláez R.,   Maté B.,  2022,
  \mn@doi [\aap] {10.1051/0004-6361/202243489}, forthcoming issue

\bibitem[\protect\citeauthoryear{Nguyen, Baouche, Congiu, Diana, Pagani  \&
  Dulieu}{Nguyen et~al.}{2018}]{Nguyen2018}
Nguyen T.,  Baouche S.,  Congiu E.,  Diana S.,  Pagani L.,   Dulieu F.,  2018,
  \mn@doi [A&A] {10.1051/0004-6361/201832774}, 619, 111

\bibitem[\protect\citeauthoryear{Noble, Congiu, Dulieu  \& Fraser}{Noble
  et~al.}{2012a}]{Noble2012}
Noble J.~A.,  Congiu E.,  Dulieu F.,   Fraser H.~J.,  2012a, \mn@doi [\mnras]
  {10.1111/j.1365-2966.2011.20351.x}, 421, 768

\bibitem[\protect\citeauthoryear{Noble, Theule, Mispelaer, Duvernay, Danger,
  Congiu, Dulieu  \& Chiavassa}{Noble et~al.}{2012b}]{Noble2012b}
Noble J.~A.,  Theule P.,  Mispelaer F.,  Duvernay F.,  Danger G.,  Congiu E.,
  Dulieu F.,   Chiavassa T.,  2012b, \mn@doi [\aap]
  {10.1051/0004-6361/201219437}, 543, A5

\bibitem[\protect\citeauthoryear{{{\"O}berg}, {van Dishoeck}, {Linnartz}  \&
  {Andersson}}{{{\"O}berg} et~al.}{2010}]{Oberg2010}
{{\"O}berg} K.~I.,  {van Dishoeck} E.~F.,  {Linnartz} H.,   {Andersson} S.,
  2010, \mn@doi [\apj] {10.1088/0004-637X/718/2/832}, \href
  {https://ui.adsabs.harvard.edu/abs/2010ApJ...718..832O} {718, 832}

\bibitem[\protect\citeauthoryear{Oganov \& Glass}{Oganov \&
  Glass}{2006}]{oganov2006crystal}
Oganov A.~R.,  Glass C.~W.,  2006, J. Chem. Phys., 124, 244704

\bibitem[\protect\citeauthoryear{Oganov, Lyakhov  \& Valle}{Oganov
  et~al.}{2011}]{oganov2011evolutionary}
Oganov A.~R.,  Lyakhov A.~O.,   Valle M.,  2011, Acc. Chem. Res., 44, 227

\bibitem[\protect\citeauthoryear{{Okoda}, {Oya}, {Abe}, {Komaki}, {Watanabe}
  \& {Yamamoto}}{{Okoda} et~al.}{2021}]{Okoda2021}
{Okoda} Y.,  {Oya} Y.,  {Abe} S.,  {Komaki} A.,  {Watanabe} Y.,   {Yamamoto}
  S.,  2021, \mn@doi [\apj] {10.3847/1538-4357/ac2c6c}, \href
  {https://ui.adsabs.harvard.edu/abs/2021ApJ...923..168O} {923, 168}

\bibitem[\protect\citeauthoryear{{Rimola}, {Skouteris}, {Balucani},
  {Ceccarelli}, {Enrique-Romero}, {Taquet}  \& {Ugliengo}}{{Rimola}
  et~al.}{2018}]{Rimola2018}
{Rimola} A.,  {Skouteris} D.,  {Balucani} N.,  {Ceccarelli} C.,
  {Enrique-Romero} J.,  {Taquet} V.,   {Ugliengo} P.,  2018, \mn@doi [ACS Earth
  Space Chem.] {10.1021/acsearthspacechem.7b00156}, \href
  {https://ui.adsabs.harvard.edu/abs/2018ECS.....2..720R} {2, 720}

\bibitem[\protect\citeauthoryear{{Scibelli} \& {Shirley}}{{Scibelli} \&
  {Shirley}}{2020}]{Scibelli2020}
{Scibelli} S.,  {Shirley} Y.,  2020, \mn@doi [\apj] {10.3847/1538-4357/ab7375},
  \href {https://ui.adsabs.harvard.edu/abs/2020ApJ...891...73S} {891, 73}

\bibitem[\protect\citeauthoryear{Shimonishi, Nakatani, Furuya  \&
  Hama}{Shimonishi et~al.}{2018}]{Shimonishi_2018}
Shimonishi T.,  Nakatani N.,  Furuya K.,   Hama T.,  2018, \mn@doi [\apj]
  {10.3847/1538-4357/aaaa6a}, 855, 27

\bibitem[\protect\citeauthoryear{{Skouteris}, {Balucani}, {Ceccarelli},
  {Vazart}, {Puzzarini}, {Barone}, {Codella}  \& {Lefloch}}{{Skouteris}
  et~al.}{2018}]{Skouteris2018}
{Skouteris} D.,  {Balucani} N.,  {Ceccarelli} C.,  {Vazart} F.,  {Puzzarini}
  C.,  {Barone} V.,  {Codella} C.,   {Lefloch} B.,  2018, \mn@doi [\apj]
  {10.3847/1538-4357/aaa41e}, \href
  {https://ui.adsabs.harvard.edu/abs/2018ApJ...854..135S} {854, 135}

\bibitem[\protect\citeauthoryear{Speedy, Debenedetti, Smith, Huang  \&
  Kay}{Speedy et~al.}{1996}]{Speedy1996}
Speedy R.~J.,  Debenedetti P.~G.,  Smith R.~S.,  Huang C.,   Kay B.~D.,  1996,
  \mn@doi [J. Chem. Phys.] {10.1063/1.471869}, 105, 240

\bibitem[\protect\citeauthoryear{Tait, Dohnálek, Campbell  \& Kay}{Tait
  et~al.}{2005}]{tait2005b}
Tait S.~L.,  Dohnálek Z.,  Campbell C.~T.,   Kay B.~D.,  2005, \mn@doi [J.
  Chem. Phys.] {10.1063/1.1883629}, 122, 164708

\bibitem[\protect\citeauthoryear{{Vastel}, {Ceccarelli}, {Lefloch}  \&
  {Bachiller}}{{Vastel} et~al.}{2014}]{vastel2014}
{Vastel} C.,  {Ceccarelli} C.,  {Lefloch} B.,   {Bachiller} R.,  2014, \mn@doi
  [\apjl] {10.1088/2041-8205/795/1/L2}, \href
  {https://ui.adsabs.harvard.edu/abs/2014ApJ...795L...2V} {795, L2}

\bibitem[\protect\citeauthoryear{{Vasyunin}, {Caselli}, {Dulieu}  \&
  {Jim{\'e}nez-Serra}}{{Vasyunin} et~al.}{2017}]{Vasyunin2017}
{Vasyunin} A.~I.,  {Caselli} P.,  {Dulieu} F.,   {Jim{\'e}nez-Serra} I.,  2017,
  \mn@doi [\apj] {10.3847/1538-4357/aa72ec}, \href
  {https://ui.adsabs.harvard.edu/abs/2017ApJ...842...33V} {842, 33}

\bibitem[\protect\citeauthoryear{{Vazart}, {Ceccarelli}, {Balucani}, {Bianchi}
  \& {Skouteris}}{{Vazart} et~al.}{2020}]{vazart2020}
{Vazart} F.,  {Ceccarelli} C.,  {Balucani} N.,  {Bianchi} E.,   {Skouteris} D.,
   2020, \mn@doi [\mnras] {10.1093/mnras/staa3060}, \href
  {https://ui.adsabs.harvard.edu/abs/2020MNRAS.499.5547V} {499, 5547}

\bibitem[\protect\citeauthoryear{Wakelam et~al.,}{Wakelam
  et~al.}{2012}]{Wakelam2012}
Wakelam V.,  et~al., 2012, \mn@doi [\apjs] {10.1088/0067-0049/199/1/21}, 199,
  21

\bibitem[\protect\citeauthoryear{{Wakelam}, {Loison}, {Mereau}  \&
  {Ruaud}}{{Wakelam} et~al.}{2017}]{Wakelam2017BE}
{Wakelam} V.,  {Loison} J.~C.,  {Mereau} R.,   {Ruaud} M.,  2017, \mn@doi [Mol.
  Astrophys.] {10.1016/j.molap.2017.01.002}, \href
  {https://ui.adsabs.harvard.edu/abs/2017MolAs...6...22W} {6, 22}

\bibitem[\protect\citeauthoryear{{Yang} et~al.,}{{Yang}
  et~al.}{2021}]{Yang2021}
{Yang} Y.-L.,  et~al., 2021, \mn@doi [\apj] {10.3847/1538-4357/abdfd6}, \href
  {https://ui.adsabs.harvard.edu/abs/2021ApJ...910...20Y} {910, 20}

\bibitem[\protect\citeauthoryear{{Zhou} et~al.,}{{Zhou}
  et~al.}{2022}]{Zhou2022}
{Zhou} C.,  et~al., 2022, arXiv e-prints, \href
  {https://ui.adsabs.harvard.edu/abs/2022arXiv220202157Z} {p. arXiv:2202.02157}

\makeatother
\end{thebibliography}








\bsp	
\label{lastpage}
\end{document}